\documentclass[aps,prc,superscriptaddress,twoside,twocolumn,nofootinbib,showpacs]{revtex4}
\usepackage{amsmath,amssymb}
\usepackage{mathtools}
\usepackage{bbold}
\usepackage[usenames, dvipsnames]{xcolor}
\usepackage{braket}
\usepackage{graphicx,bm}
\usepackage{slashed}
\usepackage{booktabs}
\usepackage{tensor}
\usepackage{multirow}
\usepackage{float}

\usepackage{kbordermatrix}% http://www.hss.caltech.edu/~kcb/TeX/kbordermatrix.sty

\usepackage{changes}

\usepackage{enumerate}

\usepackage{pifont}% http://ctan.org/pkg/pifont

\definechangesauthor[name={Yeunhwan}, color=blue]{YH}

\allowdisplaybreaks

\newcommand{\Gth}{\Gamma_{\mathrm{th}}}

\newcommand{\pth}{p_{\mathrm{th}}}
\newcommand{\ebth}{\varepsilon_{\mathrm{th}}}
\newcommand{\pcd}{p_{\mathrm{cold}}}
\newcommand{\ebcd}{\varepsilon_{\mathrm{cold}}}

\newcommand{\mbr}{\mathbf{r}}
\newcommand{\mbk}{\mathbf{k}}
\newcommand{\mbkp}{\mathbf{k}^{\prime}}

\newcommand{\taun}{\tau_n}
\newcommand{\taup}{\tau_p}

\newcommand{\al}{\alpha_L}
\newcommand{\au}{\alpha_U}

\newcommand{\bl}{\beta_L}
\newcommand{\bu}{\beta_U}

\newcommand{\el}{\eta_L}
\newcommand{\eu}{\eta_U}

\newcommand{\msun}{M_\odot}

\begin{document}

\title{Enhanced adiabatic index for hot neutron-rich matter from microscopic nuclear forces}
\date{\today}

%\author{Yeunhwan \surname{Lim} }
%\email{ylim@tamu.edu}
%\affiliation{Cyclotron Institute, Texas A\&M University, College Station, TX 77843, USA}

\author{Yeunhwan \surname{Lim} }
\email{ylim@theorie.ikp.physik.tu-darmstadt.de}
\affiliation{Max-Planck-Institut f\"ur Kernphysik, Saupfercheckweg 1, 69117 Heidelberg, Germany}
\affiliation{Institut f\"ur Kernphysik, Technische Universit\"at Darmstadt, 64289 Darmstadt, Germany}
\affiliation{ExtreMe Matter Institute EMMI, GSI Helmholtzzentrum f\"ur Schwerionenforschung GmbH, 64291 Darmstadt, Germany}

\author{Jeremy W. \surname{Holt} }
\email{holt@physics.tamu.edu}
\affiliation{Cyclotron Institute, Texas A\&M University, College Station, TX 77843, USA}
\affiliation{Department of Physics and Astronomy, Texas A\&M University, College Station, TX 77843, USA}

%% starts 5/18/2017

\begin{abstract}
We investigate the adiabatic index $\Gamma_{\mathrm{th}}$ of hot and dense nuclear matter from chiral effective field theory
and find that the results are systematically larger than from typical mean field models. We start by constructing the 
finite-temperature equation of state from chiral two- and three-nucleon forces, which we then use to fit a class of 
extended Skyrme energy density functionals. This allows for
modeling the thermal index across the full range of densities and temperatures that
may be probed in simulations of core-collapse supernovae and neutron star mergers, 
including the low-density inhomogeneous mixed
phase. For uniform matter we compare the results to analytical expressions for $\Gth$ based on Fermi 
liquid theory. The correlation between the thermal index and the effective masses 
at nuclear saturation density is studied systematically through Bayesian modeling of the nuclear equation of state.
We then study the behavior of $\Gth$ in both relativistic and non-relativistic mean field models
used in the astrophysical simulation community to complement those based on chiral effective field theory constraints
from our own study.
We derive compact parameterization formulas for $\Gth$ across the range of densities and temperatures
encountered in core collapse supernovae and binary neutron star mergers, which we suggest may be
useful for the numerical simulation community.
\end{abstract}


\maketitle

%%%%% Section 1

\section{Introduction}

Neutron stars are intriguing stellar objects whose structure and dynamics are governed by the properties 
of matter at extraordinarily high densities up to ten times that of atomic nuclei. Ongoing experimental efforts
aim to study the hot and dense matter equation of state (EOS) through medium-energy heavy-ion collisions
\cite{danielewicz02,Tsang2004,shetty07,qin12}, while astronomical observations of neutron star radii, moments of inertia, 
and tidal deformabilities
\cite{lattimer05,Steiner2005325,SLB2010,Lattimer14,abbott18} are able to probe the properties of cold
compressed matter at around twice nuclear saturation density \cite{0004-637X-550-1-426,holt19}.
In particular, the recent observation of gravitational waves from binary neutron star merger event GW170817 
\cite{abbott17,abbott18} indicates that the tidal deformability of a 1.4\,$\msun$ neutron star lies in the range
$\Lambda_{1.4}=190^{+390}_{-120}$. This measurement excludes many stiff equations of state that would give rise
to typical neutron stars with radii $R_{1.4}>13.6$\,km \cite{fattoyev18,most18,lim18a,tews18,kim18}.
Moreover, recent advances in the theoretical description of nuclear forces now enable constraints on the ground
state energy and pressure of dense nucleonic matter at and below nuclear saturation density
\cite{hebeler10a,gandolfi2012,gezerlis13,carbone14,hagen14,roggero14,wlazlowski14,Gandolfi:2015jma,sammarruca15,drischler16}. 
At higher densities theoretical uncertainties grow
rapidly \cite{holt2017a,tews18}, and the reliability of microscopic calculations becomes uncertain. 

The properties of dense nuclear matter beyond twice saturation density are highly model dependent, and not
even the appropriate degrees of freedom are known. It has been suggested that exotic matter such as hyperons
\cite{glen82,glen91,weissenborn12a,weissenborn12b,lim15h,lim18h}, 
kaon or pion condensates \cite{kaplan86,thorsson94,lee96,glen98,lim14,baym73,au74},
and strange quarks or deconfined quark matter \cite{prakash95,weber05,weissenborn11,liu19}
might exist in the inner core of neutron stars.
Imprints of this extreme matter might be confirmed through astrophysical observations
of pulsar glitches \cite{pines85,link99,andersson12,chamel13,ho15,watanabe17},
X-ray bursts \cite{guver10,SLB2010,guillot14,ozel2016,SLB2016,suleimanov16,nattil17}, 
neutron star cooling \cite{page94,yakovlev04,page2004,page2009,HH2010,page2010,Lim2017b,Lim2016zho,brown17}, 
neutron star moments of inertia \cite{morrison04,bejger05,lattimer05,lim2018i}, 
and simultaneous measurements of masses and radii,~e.g., from the NICER X-ray telescope
\cite{Bogdanov:2019owz,dimitrios14,ozel2016n}. In order to explore the widest range of high-density parameterizations 
for the cold dense matter equation of state (including possible phase transitions), theoretical modeling has employed
piecewise polytropes \cite{read09,SLB2010,hebeler10,raithel16}, the spectral representation \cite{lindblom10}, polynomial 
expansions in powers of the density or Fermi momentum \cite{fattoyev14,margueron18a,lim18a}, and speed of 
sound parameterizations \cite{chamel13,bedaque14,tews18}. 

Dynamical simulations of core-collapse supernovae
\cite{Janka96,Rampp02,Kifonidis03,Buras05,kotake06,Janka06,ott09,murphy09,roberts16} 
and neutron star mergers
\cite{Ruffert98,Ruffert99,Ruffert01,Oechslin07,bauswein10,bauswein12,korobkin12,tanaka13,Kasen2017,Thielemann17}
require also the equation of state at finite temperatures up to $50-100$\,MeV. 
Such simulations are crucial for interpreting observable electromagnetic, neutrino, and gravitational wave 
emissions as well as understanding the origin of many heavy elements through r-process nucleosynthesis 
\cite{lattimer74,lattimer76,Kasen2017,Thielemann17,radice18}.
Presently there exist a number of EOS tabulations derived from Skyrme mean field phenomenology
\cite{lattimer85,LSEOS,lim12,schneider17} and
relativistic mean field models \cite{shen98,gshen11,shen11,hempel10,hempel12,hurusawa11} appropriate
for astrophysical simulations. To explore an even wider range of parameterizations, polytropic equations of 
state have been coupled with an ideal gas ansatz for the thermal contribution
to the pressure
\begin{equation}
p_{\rm th} = (\Gth - 1) \varepsilon_{\rm th},
\end{equation} 
where $\varepsilon_{\rm th}$ is the internal energy density and $\Gth$ is the so-called adiabatic index.
The resulting analytical equations of state have numerical 
advantages over tabulations and allow for a more thorough exploration of the correlations 
among bulk neutron star properties, features of gravitational wave signals, and properties of the 
equation of state. In particular, thermal effects have been shown to modify the dominant peak 
frequency and intensity of the post-merger gravitational wave signal \cite{bauswein10,bauswein12} 
and lead to a delay in the remnant collapse time to a black hole \cite{baiotti08,bauswein10}.

\begin{figure}[t]
	\includegraphics[scale=0.42]{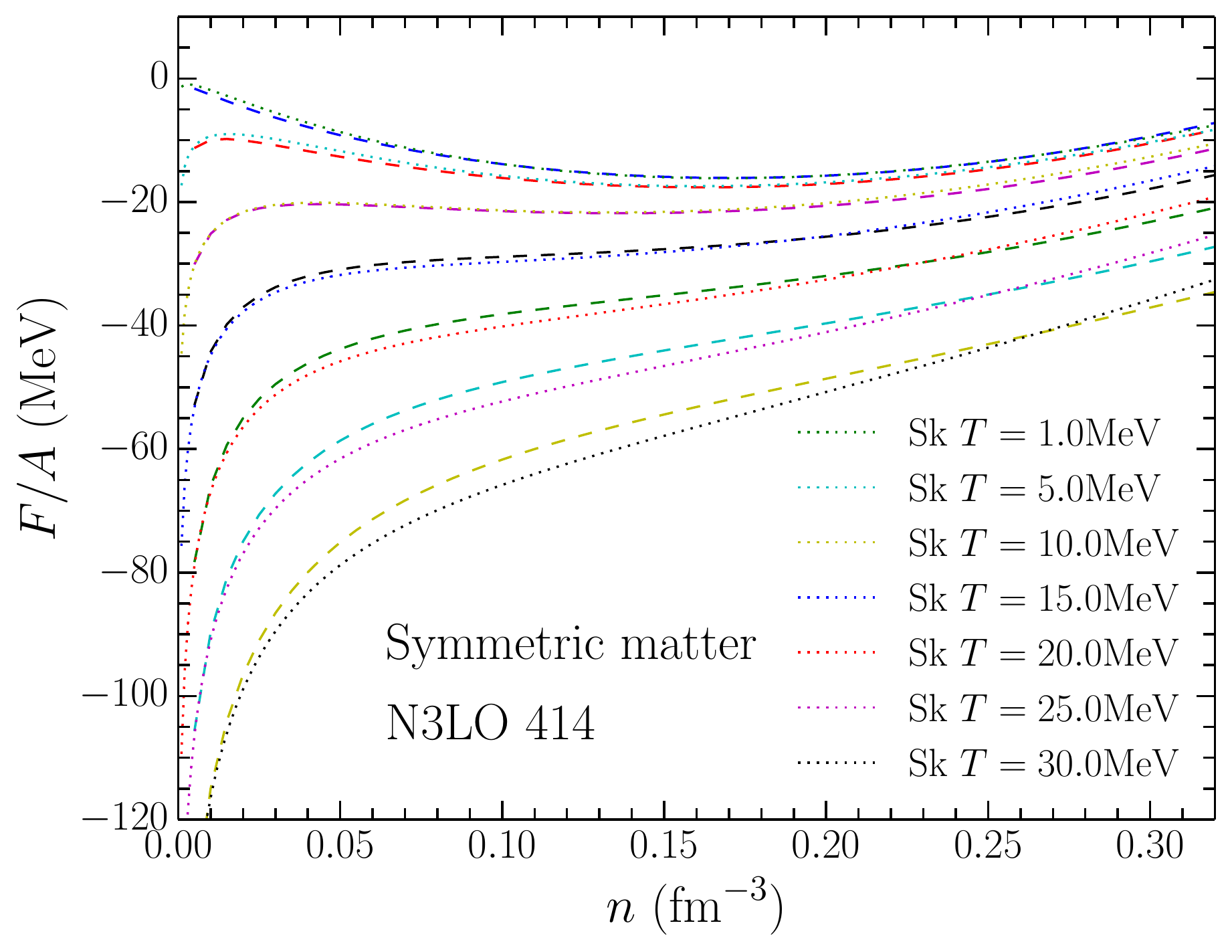}
	\caption{Free energy per baryon for isospin-symmetric matter as a function of density and temperature 
	computed from the N3LO414 chiral nuclear potential (dashed lines) and the fitted extended Skyrme
	functional (dotted lines).}
	\label{fig:frees}
\end{figure}

Previous studies \cite{baiotti08,bauswein10,bauswein12} have employed a constant value of the thermal index, 
and in the present work our aim is to provide a general parameterization for $\Gth$ in terms of the density and
temperature based on finite-temperature calculations of the equation 
of state \cite{wellenhofer14,wellenhofer15} from microscopic chiral effective field theory ($\chi$EFT).
Recent works have also studied the thermal index from microscopic many-body theory \cite{carbone19a,lu19}
and parameterized equations of state \cite{raithel19}. We derive empirical formulas for 
$\Gth$ as a function of temperature and
baryon number density for a proton fraction fixed to that of beta equilibrium matter at $T=0$\,MeV. 
For this end, we collect EOSs available in
the supernova community and build our own EOSs based on $\chi$EFT results for the free energy density. 
We present fitting functions and covariance matrices that will allow for the generation of realistic 
behaviors for $\Gth$ as a function of density and temperature within the modeled
statistical uncertainties.

The paper is organized as follows. In Section\,\ref{sec:model}, we briefly explain the nuclear force models
used for constructing the hot and dense matter EOS. In Section \ref{sec:thermal}, we present results for $\Gth$ 
from our current calculations based on chiral effective field theory and existing mean field model EOSs. We
then derive empirical formulas and covariance matrices for the description of $\Gth$ as a function of
density and temperature. In section\,\ref{sec:sum}, we summarize our results and discuss 
future directions for modeling the hot and dense matter EOS from realistic nuclear force models.

\section{Nuclear Force Models}
\label{sec:model}

\begin{figure}[t]
	\includegraphics[scale=0.42]{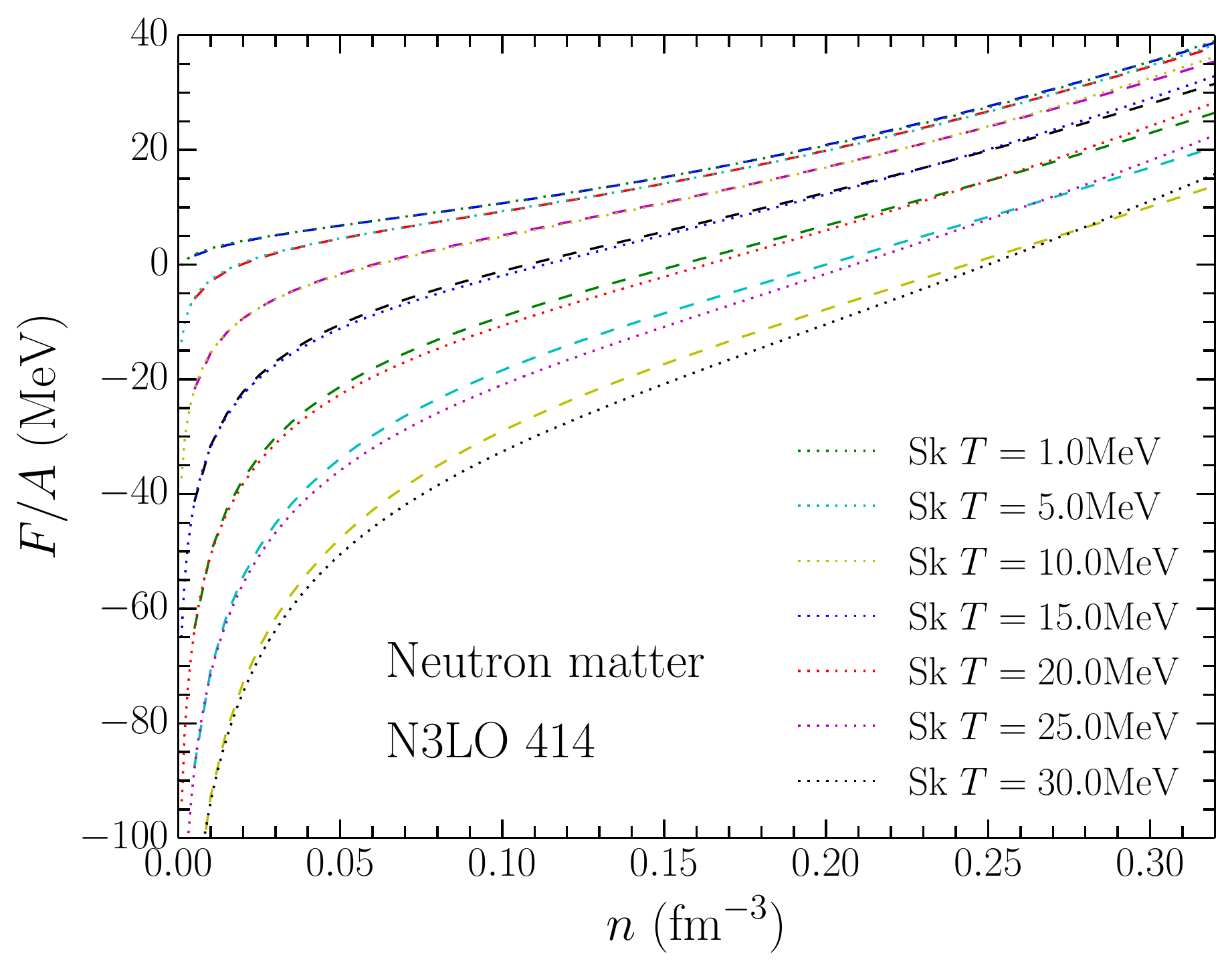}
	\caption{Free energy per baryon for pure neutron matter as a function of density and temperature 
	computed from the N3LO414 chiral nuclear potential (dashed lines) and the fitted extended Skyrme
	functional (dotted lines).}
	\label{fig:freen}
\end{figure}

Nuclear forces derived from chiral effective field theory ($\chi$EFT) are nowadays
widely used to study many properties of light and medium-mass nuclei
as well as dense uniform matter \cite{navratil07,epelbaum14,hebeler15,holt16}.
Recently, chiral nuclear forces have been used to build finite-temperature equations of state
based on many-body perturbation theory in the Matsubara imaginary time formalism 
\cite{wellenhofer14,wellenhofer15,carbone18,du19}.
However, microscopic calculations break down for the low-density inhomogeneous mixed phase 
where the spinodal instability is active
as well as the high-density phase where chiral effective field theory is no longer applicable.
We therefore choose to sample representative calculations of the hot and dense matter
equation of state from $\chi$EFT and fit the results to the form of a Skyrme-like potential model.
In particular, we employ six different chiral nuclear forces \cite{sammarruca15}: N2LO450, N2LO500, N3LO414, N3LO450,
and N3LO500, corresponding to a next-to-next-to-leading-order (N2LO) chiral potential with momentum-space
cutoff $\Lambda_\chi = 450$\,MeV, an N2LO chiral potential with momentum-space cutoff 
$\Lambda_\chi = 500$\,MeV, etc. As shown in Figs.\ \ref{fig:frees} and \ref{fig:freen}, we fit the 
microscopic results up to $n = 2n_0$ and $T<30$\,MeV. In all cases we verify that the fitted Skyrme potential 
model satisfies causality and can produce a neutron star with maximum mass $M_{\rm max} \ge 1.97\,\msun$.
For this reason, we use the extended Skyrme force model
as suggested in Ref.\ \cite{Chamel09}, where the two-nucleon interaction has the form,
\begin{equation}
\begin{aligned}
v_{ij}^{\prime} = & v_{ij} 
+ \frac{t_4}{6}(1+x_4P\sigma)n^{\epsilon_2}(\mbr) \delta(\mbr_{ij}) \\
& + \frac{1}{2}t_5 (1+x_5P_\sigma)[\mbk_{ij}^{\prime 2} n^{\gamma_1}(\mbr)\delta(\mbr_{ij})
+\delta(\mbr_{ij}) n^{\gamma_1}(\mbr)\mbk^2] \\
& + t_6(1+x_6P_\sigma)\mbkp_{ij}\cdot n^{\gamma_2}(\mbr)\mbk_{ij}\delta(\mbr_{ij})\,.
\end{aligned}
\end{equation}
This type of extension gives the energy density of uniform nuclear matter, i.e.\
without spin-orbit interaction and density gradient contributions, as
\begin{equation}
\begin{aligned}
\mathcal{E} = & \frac{1}{2M_N}(\tau_n + \tau_p)  \\ 
& + (n_n^2 + n_p^2)f_L(n) + 2 n_n n_p f_U(n)   \\
& + (n_n\tau_n + n_p\tau_p)g_L(n) + (n_n\taup + n_p\taun)g_U(n) \,,
\end{aligned}
\end{equation}
where $M_N$ is the nucleon mass, and the number density ($n$) and kinetic density ($\tau$) are defined as
\begin{equation}
n_t = \frac{1}{\pi^2} \int \frac{k^2\,dk}{1 + e^{(\epsilon_t -\mu_t)/T}}\,,\quad
\tau_t = \frac{1}{\pi^2} \int \frac{k^4\,dk}{1 + e^{(\epsilon_t -\mu_t)/T}}\,
\end{equation}
with $\mu_t$ the chemical potential of species $t=p,n$.
Here $\epsilon_t$ is the single particle energy given by
\begin{equation}
\epsilon_t(k) = \frac{k^2}{2M_t^*} + V_t \,, 
\quad \frac{1}{2M_t^*} = \frac{\delta \mathcal{E} }{\delta \tau_t}\,,
\quad V_t = \frac{\delta \mathcal{E} }{\delta n_t}\,.
\end{equation}
The functionals $f_{L,U}(n)$ and $g_{L,U}(n)$ are given as
\begin{equation}
\begin{aligned}
f_{L,U}(n) & = \alpha_{L,U} + \eta_{L,U}n^{\epsilon_1} + \lambda_{L,U} n^{\epsilon_2} \,, \\
g_{L,U}(n) & = \beta_{L,U} + \zeta_{L,U}n^{\gamma_1} + \sigma_{L,U} n^{\gamma_2} \,.
\end{aligned}
\label{lu}
\end{equation}
Note that the conventional Skyrme force model contains only
$\al$, $\au$, $\bl$, $\bu$, and $\el$, $\eu$. 
However, we use the extended model to obtain a better description of the free energy 
computed at finite temperature from chiral nuclear potentials.

\begin{figure}[t]
	\includegraphics[scale=0.55]{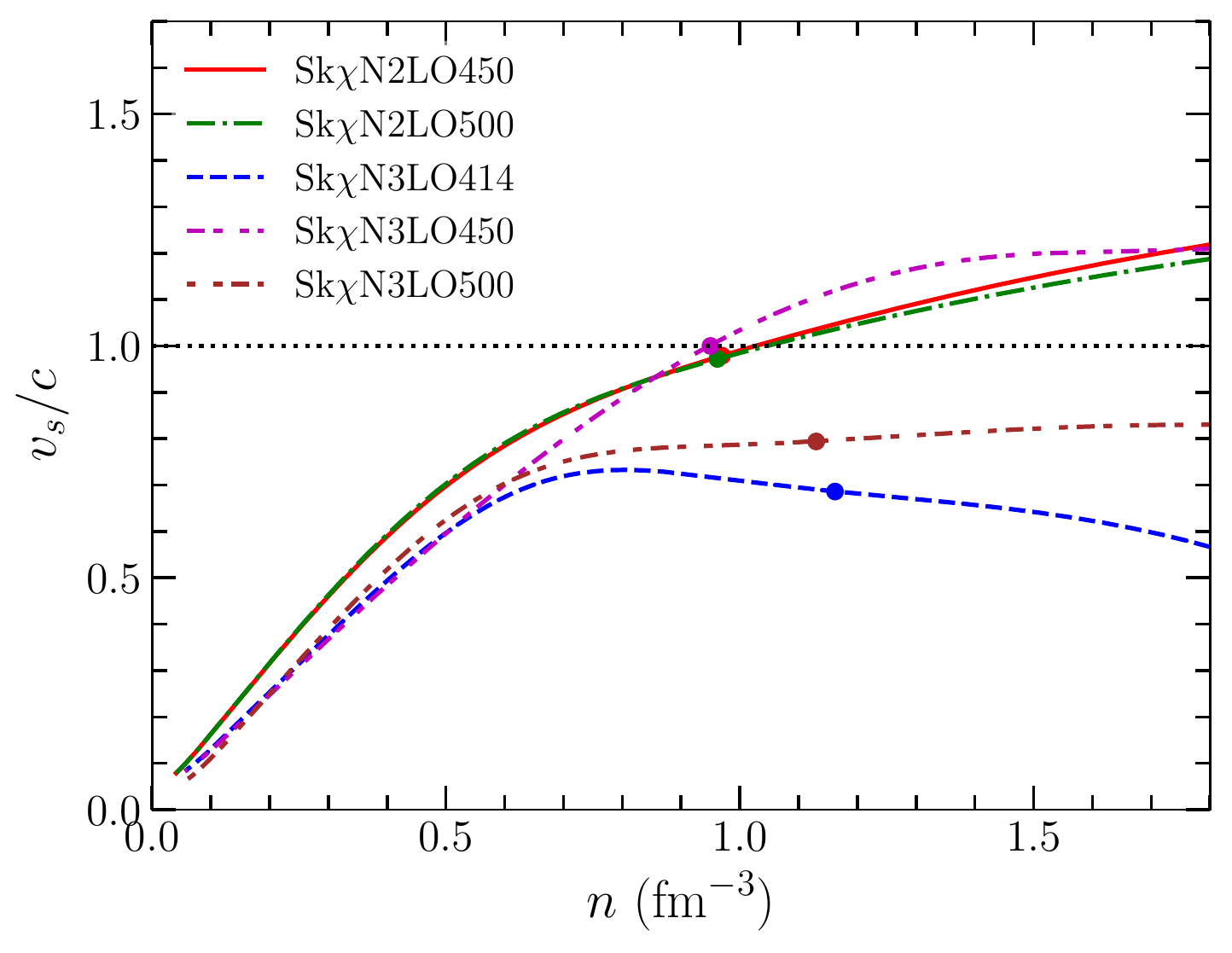}
	\caption{The speed of sound as a function of density for beta-equilibrated nuclear matter from the 
	five Skyrme energy density functionals constructed in the present work. The dots denote the density of the
	maximum neutron star mass for each associated equation of state.}
	\label{fig:speed}
\end{figure}

The effective mass in the extended form is expressed as
\begin{equation}\label{eq:effm}
\frac{1}{2M_N^*}   =  \frac{1}{2M_N} +  g_L(n) n_n +  g_U(n) n_p\,.
\end{equation}
In contrast to effective masses computed from microscopic nuclear forces, the effective mass
from the Skyrme formalism does not depend explicitly on the temperature. In particular,
thermal effects are known \cite{donati94} to strongly suppress the energy dependence of the 
nucleon single-particle potential, thereby reducing the normally strong enhancement of the effective mass close to
the Fermi surface \cite{brown63,bertsch68}. At high temperatures we therefore expect that
the momentum dependence of the nucleon single-particle potential is primarily responsible for the effective 
mass, which would exhibit a behavior more similar to that given in Eq.\ \eqref{eq:effm}. However,
mean field models may not be well suited to reproduce these higher-order temperature effects.
When we set the power of the density in the extra effective mass contribution $\gamma_1 =\gamma_2$, we then
have ten parameters to be determined for fixed $\epsilon_1(=1/3)$, $\epsilon_2(=1)$ 
and $\gamma (=2/3 ~\text{or}~ 1)$. In this work, we use the ten parameters to fit 
the free energy per baryon from chiral effective field theory calculations at finite temperature
as well as zero temperature.

In Fig.\ \ref{fig:frees} and \ref{fig:freen} we show the free energy per baryon of symmetric nuclear
matter and pure neutron matter as a function of density and temperature from $\chi$EFT (dashed curves)
and the corresponding non-relativistic model calculation (dotted curves). We see that 
as the temperature increases, it becomes more difficult to fit the chiral EFT results to the form of a 
Skyrme energy density functional, as observed also in Ref.\ 
\cite{PhysRevC.93.065801}.
Note that the Skyrme force model includes only the first-order Hartree-Fock 
contribution to the free energy. In contrast, the results from chiral effective
field theory include second-order contributions plus Hartree-Fock self-energies for the intermediate
states as described in Ref.\ \cite{wellenhofer14}.

In Fig.\ \ref{fig:speed} we show the speed of sound in beta stable nuclear matter
at $T=0$\,MeV for the five Skyrme interactions fitted in the present work to the hot and dense matter equations
of state from chiral effective field theory. 
We note that some of the models become acausal around $n\simeq 1.0\,\mathrm{fm}^{-3}$,
but these densities typically lie beyond the maximal density reached at the center of the maximum
mass neutron star. Acausality in non-relativistic models may be remedied in a thermodynamically
consistent manner according to the general formalism in Ref.\ \cite{constantinou17} (see also Ref.\ \cite{du19}).
In the present work we consider only equations of state that give rise to subluminal speeds of sound at all
densities within the neutron stars considered.

%Alternatively, remedied by adding extra density functional or setting a critical density to replace the original energy density functional. That is, we can set the energy density and pressure of the total baryon number density $n>n_h$,
%\begin{equation}
%\label{eq:neweos}
%\begin{aligned}
%\varepsilon(n) & = \varepsilon_h \left(  \frac{n}{n_h} \right)^{1 + \frac{p_h}{\varepsilon_h}} \,, \\
%p(n) & = p_h \left(  \frac{n}{n_h} \right)^{1 + \frac{p_h}{\varepsilon_h}}\,,
%\end{aligned}
%\end{equation}
%where $\varepsilon_h$ and $p_h$ is the energy density and pressure at $n=n_h$. 
%The new set of equation above is continuous over the energy density and pressure.
%We can choose 
%$n_h$ where the chiral effective field theory is not valid or the densities
%where the speed of sound becomes conformal limit. 
%Note that the Eq.\,\eqref{eq:neweos} gives the speed of sound,
%\begin{equation}
%c_s^2 = \frac{\partial p}{\partial \varepsilon} = \frac{dp/dn}{d\varepsilon/dn} = \frac{p_h}{e_h}\,.
%\end{equation}
%If the new EOS is not stiff enough, we may 
%employ several $n_h$'s, we can make any EOSs to stiff enough to achieve 2.0\,$\msun$ neutron stars.
%In this work, however, we don't use this type piecewise EOS since it doesn't provide the
%information of particle fraction of neutrons and protons in ground state.\\

\begin{figure}[t]
	\includegraphics[scale=0.55]{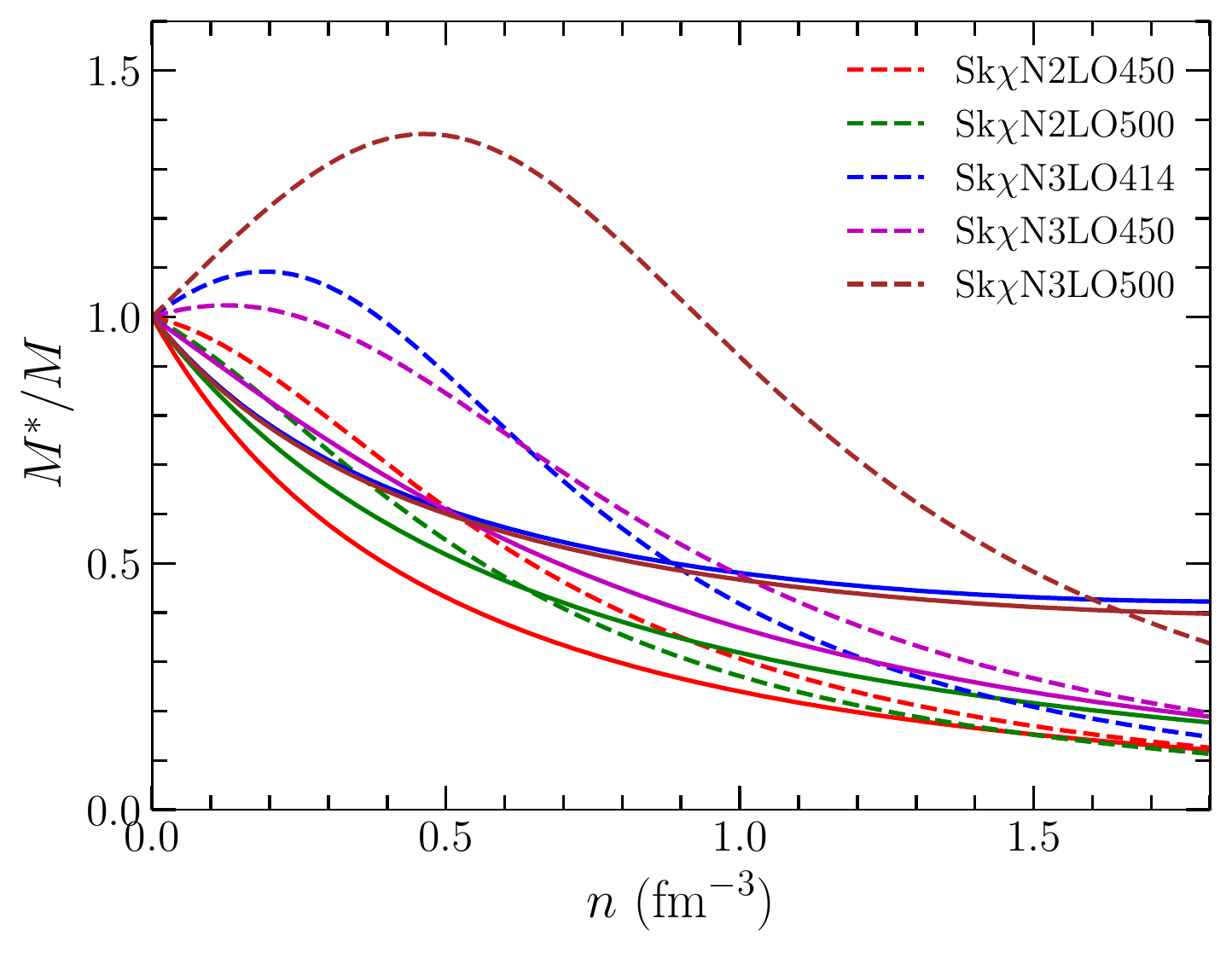}
	\caption{The neutron effective mass in pure neutron matter (dashed lines) and the nucleon effective 
	mass in symmetric nuclear matter (solid lines) as a function of density for each Skyrme effective 
	interaction constructed in the present work.}
	\label{fig:effm}
\end{figure}

In Fig.\ \ref{fig:effm} we plot the nucleon effective mass in both symmetric nuclear matter (solid line)
and pure neutron matter (dashed lines) for each of the Skyrme effective interactions constructed in the
present work. The effective mass is obtained by employing Eq.\,\eqref{eq:effm}
and fitting the free energy density of nuclear matter computed in chiral effective field theory as
described above. We see that at nuclear matter saturation density, the neutron effective mass in pure neutron 
matter takes on the range of values $0.86<M^*/M<1.19$, which is consistent with other phenomenological and 
microscopic predictions \cite{li18,Ismail19}. In symmetric nuclear matter at saturation density, the nucleon effective mass 
lies in the more narrow range $0.74<M^*/M<0.88$, also consistent with previous theoretical
modeling \cite{holt12npa,li18}.

\begin{figure}[t]
	\includegraphics[scale=0.55]{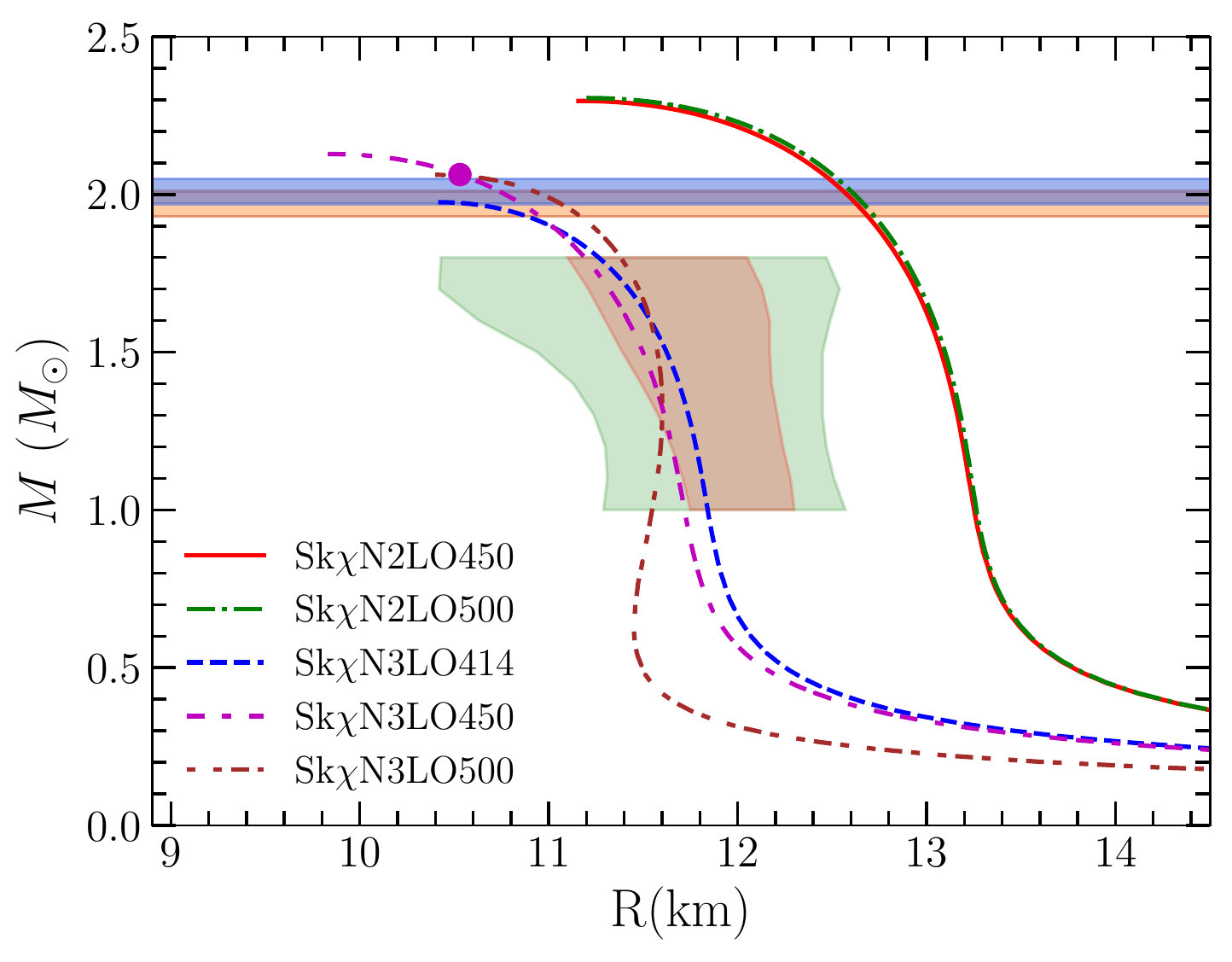}
	\caption{The neutron star mass-radius relationship for each of the Skyrme effective interactions
	constructed in the present work. The dot denotes the location at which the Sk$\chi$N3LO450 interaction gives a
	speed of sound that exceeds the speed of light.}
	\label{fig:nsmr}
\end{figure}

To test the predictions of the different EOS parameterizations, we calculate in Fig.\ \ref{fig:nsmr}
the individual mass-radius relations for cold neutron stars. We see that in all cases the equations of state
produce neutron stars satisfying the maximum mass constraint $M_{\rm max} \gtrsim 1.97\,\msun$ from 
Refs.\ \cite{demorest10,antoniadis13}. The Skyrme effective interaction fitted to 
reproduce the equation of state from the N3LO450 chiral nuclear potential is the only one for
which the speed of sound becomes superluminal before the maximum neutron star mass is reached.
Nevertheless, the reduced maximum mass that is consistent with causality still exceeds $1.97\,\msun$.
In our analysis below, we consider only neutron stars for which the speed of sound is subluminal, and
therefore we remove the most massive neutron stars produced by the Skyrme N3LO450 interaction.
In computing the mass-radius relationships, we have included realistic modeling of the neutron star outer
and inner crusts \cite{lim2017a}. This occurs when the matter density is below roughly half saturation density and forms
nuclear clusters. The composition is therefore ionized heavy nuclei in the inner and outer crust of neutron 
stars with uniform nuclear matter in the outer core \cite{BBP,BPS,Chamel2008,hebeler10,hebeler13,lim2017a}.

After determining the Skyrme model parameters, we use the liquid drop model technique to
construct the hot dense matter EOS.
In the liquid drop model, a single heavy nucleus exists in the Wigner-Seitz cell
surrounded by a gas of neutrons, protons, electrons, and alpha particles.
The free energy density is given by \cite{LSEOS,lim12},
\begin{equation}
\begin{aligned}
F(n,x,T) & = F_{\text{dense}} + F_{\text{dilute}} 
          + F_{\text{surface}} + F_{\text{Coulomb}}\\
         & + F_{\text{trans}} + F_\alpha + F_e + F_\gamma,
\end{aligned}
\end{equation}
where $F_{\text{dense}}$ is the bulk free energy density from the heavy nucleus in
the Wigner Seitz cell, $F_{\text{dilute}} $ is the free energy density 
from nucleons outside of the heavy nucleus,
$F_{\text{surface}}$ is the surface energy density of the heavy nucleus,
$F_{\text{Coulomb}}$ is the Coulomb energy density between proton-proton,
proton-electron, and electron-electron systems,
$F_{\text{trans}}$ is the translational energy density of the heavy nucleus,
$ F_\alpha$ is the alpha particle free energy density,
$ F_e$ is the electron (and positron) contribution,
and $F_\gamma$ is the photon contribution.
We then minimize $F(n,x,T)$ as a function of density, proton fraction and temperature with respect to the
dependent variables 
$(u,n_i,x_i,n_o,x_o,n_\alpha)$ with the two constraints (i) baryon number conservation and (ii) charge neutrality:
\begin{equation}
\begin{aligned}
n & = u n_i + (1-u)[4n_\alpha + (1-n_\alpha v_\alpha)n_o], \\
nx & = u n_ix_i + (1-u)[2n_\alpha + (1-n_\alpha v_\alpha)n_ox_o], \\
\end{aligned}
\end{equation}
where $n_i$ is the baryon number density of the heavy nucleus,
$x_i$ is the proton fraction of the heavy nucleus,
$n_o$ is the baryon number density of nucleons outside the heavy nucleus,
$x_o$ is the proton fraction of nucleons outside the heavy nucleus,
$n_\alpha$ is the alpha particle density, and $v_\alpha$ is the volume of alpha particle ($v_\alpha = 24\,\mathrm{fm}^3 $).

%%%%%%%%%%%%%%%%%%%%%%%%%%%%%%%%%%%%%%%

\begin{figure*}[t]
  \includegraphics[scale=0.6]{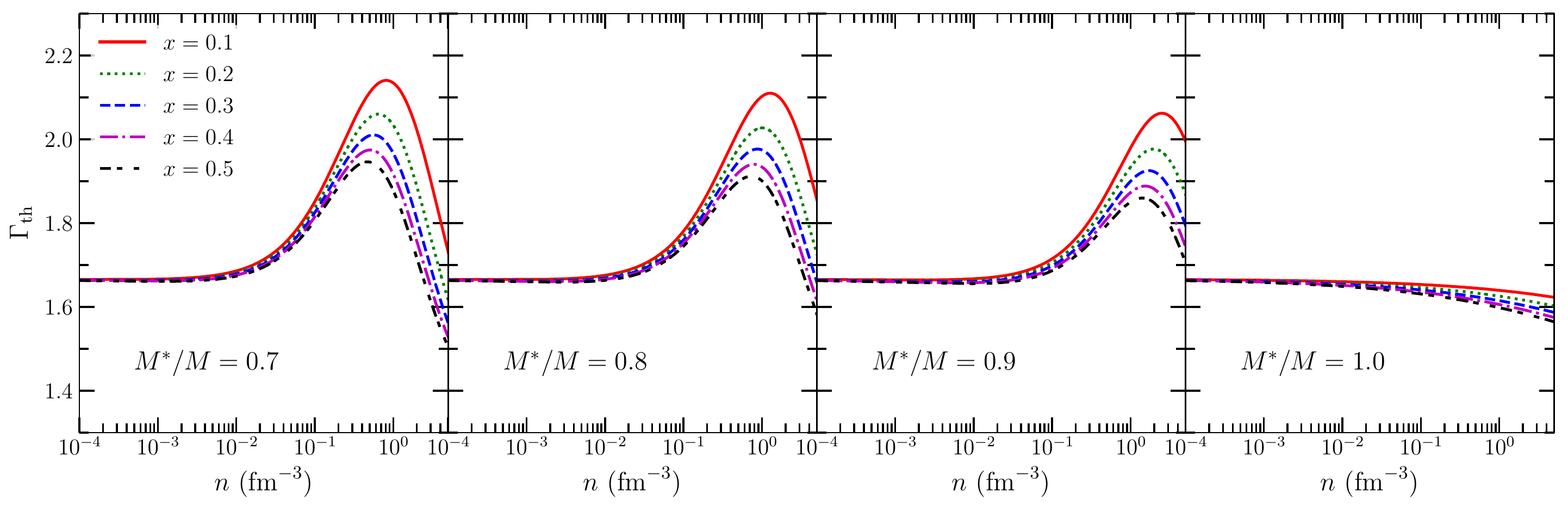}
  \caption{ 
  The thermal index $\Gth$ as a function of density $n$ and proton fraction $x$ calculated in the Fermi liquid
  theory quasiparticle approximation from Eqs.\ (\ref{gamt}) and (\ref{eq:gth1}).
  The nucleon effective mass as a function of density is assumed to follow the standard Skyme form, and the value at saturation
   density is labeled as $M^*/M$ in the different panels.}
  \label{fig:fltgth}
\end{figure*}

\section{Thermal Index: $\Gamma_{\mathrm{th}}$}
\label{sec:thermal}
The pressure of nuclear matter can be written as the sum of a
zero-temperature contribution and a thermal contribution:
\begin{equation}
p = p_{\mathrm{cold}} + p_{\mathrm{th}}\,,
\end{equation}
where all quantities are defined at the same density and proton fraction. 
Since there are large uncertainties in the nuclear matter equation of state at high densities, 
it may be useful to first construct the pressure of cold nuclear matter using piecewise polytropes.
Theoretical modeling of the thermal contribution $p_{\mathrm{th}}$ would then allow for the extension 
of the equation of state to finite temperature. In the present work we will employ a number of realistic 
equations of state at non-zero temperature to extract the thermal contribution and develop a useful
parameterization with (correlated) uncertainty estimates on the model parameters.

It is natural to formulate the thermal pressure in terms of an adiabatic index 
$\Gth$ defined as
\begin{equation}
\Gth = \frac{\pth}{\ebth} +1\,,
\label{gamt}
\end{equation}
where $\pth$ and $\ebth$ are the pressure and
internal energy density contributions from finite temperature respectively:
\begin{equation}
p_{\mathrm{total}} = \pth + \pcd\,; \quad
\varepsilon_{\mathrm{total}} = \ebth + \ebcd\,.
\end{equation}
We note that $\Gth$ is a 
function of the independent variables used for constructing the nuclear EOS.
In previous simulations of core-collapse supernovae and neutron star mergers \cite{Janka92,bauswein10},  
the most widely used values for $\Gth$ are $1.5$ or $2.0$, which may be obtained by averaging 
the true adiabatic index from full finite-temperature equations of state
over the typical range of temperatures, densities and proton fractions encountered in simulations. 
Since $\Gth$ can vary significantly as a function of density and temperature \cite{bauswein10}, 
the constant $\Gth$ approximation should be replaced with more realistic modeling.

Fermi liquid theory (FLT) provides a clear conceptual framework based on the quasiparticle approximation to understand
the thermal excitations of Fermi systems at low temperatures ($T \ll k_F^2 / 2M$), and in particular the adiabatic index $\Gth$. 
Since the quasiparticle effective mass
is directly related to the density of states, it plays a key role in entropy generation and other thermal
properties of dense matter. In particular, higher effective masses are associated with a larger density of 
states and therefore a reduced thermal pressure, leading for instance to faster contraction of a newly-born
proto-neutron star following stellar core collapse \cite{Yasin18}.
From Fermi liquid theory, the thermal contribution to the internal energy density
and pressure of nuclear matter
composed of protons, neutrons, and electrons is given by \cite{Constantinou15}
\begin{equation}\label{eq:gth1}
\begin{aligned}
\ebth & = \sum_{i=n,p,e}  n_i a_i T^2\,,\\
\pth & = \frac{2}{3}T^2 \sum_{i=n,p,e} a_i n_i 
\left[ 1 - \frac{3n_i}{2M_i^*}\frac{\partial M_i^*}{\partial n_i} \right]\,,
\end{aligned}
\end{equation}
where $a = \frac{\pi^2 M^*}{2p_F^2}$ is the level density parameter at the Fermi surface,
and the effective mass for relativistic electrons is given as
$m_e^* = \sqrt{p_{F_e}^2 + m_e^2}$.
Note that for a Fermi gas with fixed composition and temperature-independent effective mass, 
the FLT approximation does not exhibit a 
temperature dependence in $\Gth$.

For orientation, in Fig.\ \ref{fig:fltgth} we show the parametric dependence of the thermal index $\Gth$ on the
effective mass for different values of the proton fraction\,($x$) and the baryon number density\,($n$) using the 
Fermi liquid theory expressions in Eqs.\ (\ref{gamt}) and (\ref{eq:gth1}).
For the moment, we assume for simplicity that the nucleon effective mass $M^*$ at the Fermi surface does not vary with
the isospin asymmetry and depends on the baryon number density as in typical Skyrme models:
\begin{equation}\label{eq:effm1}
M^* = \frac{M}{1 + \beta n}\,.
\end{equation}
Note that in our extended Skyrme parametrization, Eq.\ \eqref{eq:effm}, the effective mass
can have a more complicated density dependence. 
In Fig.\ \ref{fig:fltgth} the effective masses are labeled by their values at nuclear matter saturation density $n_0$.
We see that the behavior of $\Gth$ above about half saturation density is strongly sensitive to
the nucleon effective mass. In particular, a lower effective mass naturally gives rise to higher values of
$\Gth$ at high density regions.

\section{Results}

We now consider two different methods for parameterizing the finite-temperature equation of state
based on microscopic many-body calculations. In the first approach we again take a parametrized form
for the effective mass, which we then combine with a wide range of cold neutron star equations of state 
constrained by nuclear theory and experiment \cite{lim18a}. The limitation is that at low density and 
finite temperature we assume 
the presence of uniform matter, since constructing an ensemble of nuclei in a gas of unbound protons 
and neutrons requires additional modeling beyond the level of Fermi liquid theory, e.g., the nucleus volume fraction 
in the Wigner-Seitz cell approximation. In the second approach we consider
full finite-temperature equations of state based on mean field models fitted to results from chiral effective field theory. 
This allows for a reliable treatment of the inhomogeneous mixed phase at low
density and finite temperature. The comparison of the two methods allows us to
explore the effects of nuclear clusters on the adiabatic index.

The first approach is based on the Bayesian 
statistical framework developed in Ref.\ \cite{lim18a} in which constraints from chiral effective 
field theory define the prior probability distributions for the parameters $\vec a$ and $\vec b$ entering
in a power series expansion of the zero-temperature equation of state:
\begin{equation}
\label{eq:exp}
\begin{aligned}
\frac{E}{A}(k_F,x=0.5) &= 2^{2/3}\frac{3}{5}\frac{k_F^2}{2M} + \frac{k_F^3}{9\pi^2} \sum_{i=0}^{3} \frac{a_i}{i!}\, \beta^i \\
\frac{E}{A}(k_F,x=0) & = \frac{3}{5}\frac{k_F^2}{2M} + \frac{k_F^3}{9\pi^2} \sum_{i=0}^{3} \frac{b_i}{i!}\, \beta^i\,,
\end{aligned}
\end{equation}
where $E/A$ is the energy per particle, $\beta \equiv (k_F - k_F^r) / k_F^r$ with $k_F^r$ an arbitrary reference Fermi momentum,
and in both expansions we define $k_F = ( 3 \pi ^ 2 n ) ^ {1/3}$. In practice, this is achieved by fitting the 
equations of state from chiral effective field theory to the form given in Eq.\ \eqref{eq:exp} from which we extract the full
covariance matrices for the $\vec a$ and $\vec b$ parameters. Theoretical uncertainties are 
estimated by varying (i) the order in the chiral expansion, (ii) the order in the many-body perturbation theory calculation
of the equation of state, and (iii) the momentum-space cutoff in the chiral interaction that sets the resolution scale in 
coordinate space. Experimental data from medium-mass and heavy
nuclei (e.g., the saturation energy $B$ and density $n_0$ of symmetric nuclear matter, 
the nuclear incompressibility $K$, and symmetry energy $S_2$) 
are then incorporated into Bayesian likelihood functions for the $\vec a$ and $\vec b$ parameters. This framework
allows us to sample a very large set of cold dense matter equations of state for isospin-symmetric nuclear matter
and pure neutron matter.

For intermediate values of the proton fraction $x = Z/A$, we expand the free energy per particle $F/A$ in a Taylor 
series around isospin-symmetric nuclear matter at saturation density:
\begin{eqnarray}
\label{eq:frees}
\frac{F}{A}(n,x,T) & = & \frac{E}{A}(n,x=0.5,T=0) + S_2(n)(1-2x)^2 \nonumber \\ 
&&- \frac{T^2}{2n}\left( \frac{\pi}{3} \right)^{2/3} [M_n^* n_n^{1/3} + M_p^* n_p^{1/3}] \,,
\end{eqnarray}
where $S_2(n)$ is the density-dependent symmetry energy. In the present work, the Bayesian likelihood function 
associated with the symmetry energy $S_2(n)$ is 
constructed from empirical constraints \cite{Lattimer:2012xj} on the symmetry energy at saturation density $J$ together with universal 
correlations among $J$, the symmetry energy slope parameter $L$, and symmetry incompressibility $K_{\text{sym}}$ 
\cite{holt18}.
The proton fraction in beta equilibrium matter can be determined from $S_2(n)$ according to
\begin{equation}
\begin{aligned}
\mu_n - \mu_p  & =  -\frac{\partial f}{\partial x}  = \mu_e \\
\implies 4 S_2(n)(1-2x) & = (3\pi^2 n x)^{1/3}\,.
\end{aligned}
\end{equation}
The last term in Eq.\,\eqref{eq:frees} can be derived in the quasiparticle approximation from Landau Fermi liquid theory 
at finite temperature.

\begin{figure*}
\centering
\includegraphics[scale=0.49]{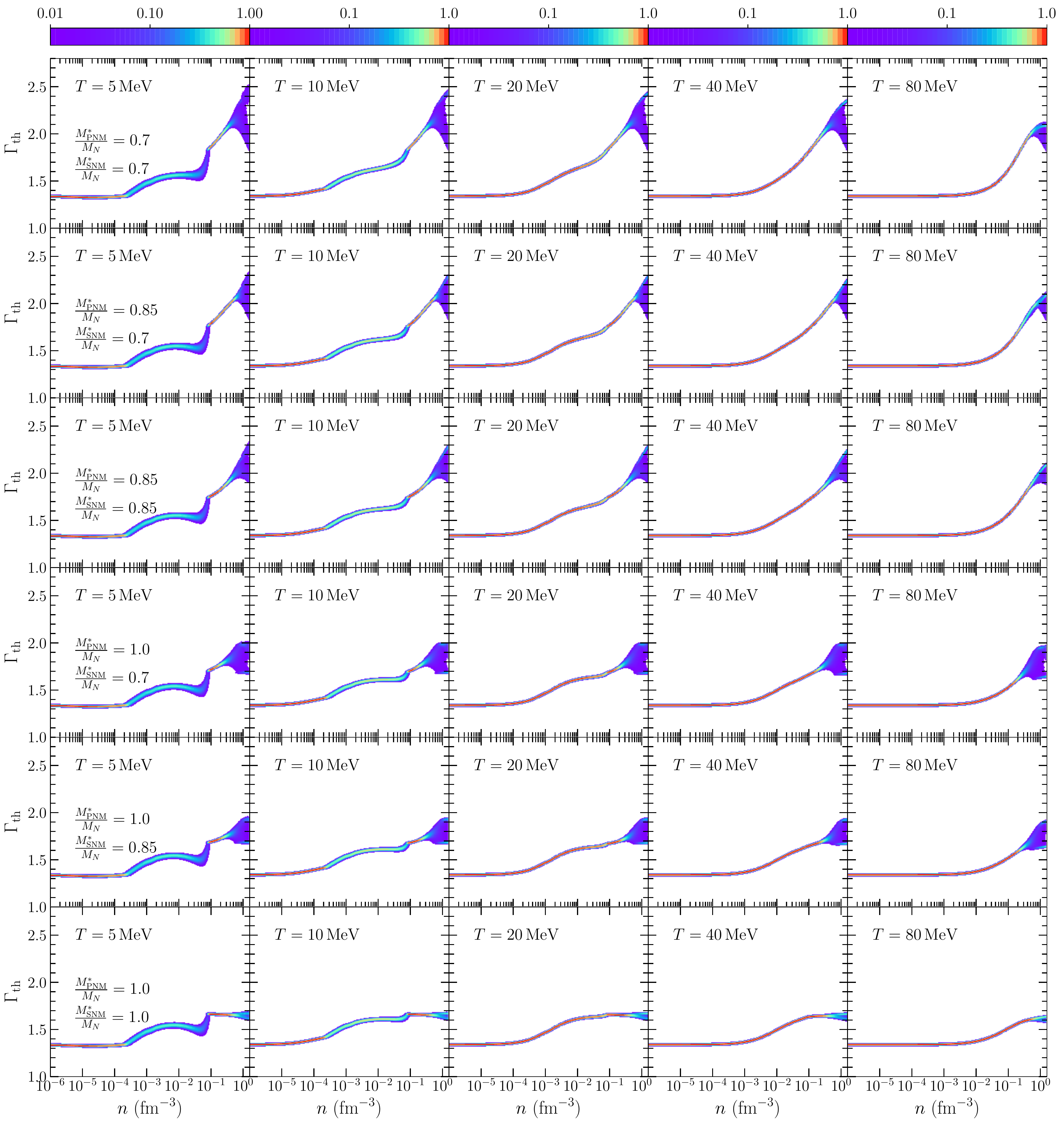}
\caption{
Contour plots for the adiabatic index $\Gth$ in the Fermi liquid approximation given by Eq.\ (\ref{eq:gth1}) for
different values of the nucleon effective mass in symmetric nuclear matter $M^*_{SNM}/M_N$
and pure neutron matter $M^*_{PNM}/M_N$ (labeled by their values at saturation density).
The matter composition is obtained by enforcing beta equilibrium on the cold dense matter equation of state 
sampled within the Bayesian statistical modeling of Ref.\ \cite{lim18a}.}
\label{fig:gthdist}
\end{figure*}

We note that Eq.\,\eqref{eq:gth1} does not necessarily describe the thermal contributions to the energy density and pressure
at low density, where nuclear matter is clustered.
In particular, there is no neutron or proton gas when the density is lower than the neutron drip density 
$n_{\text{drip}}=2.4\times 10^{-4}\,\mathrm{fm}^{-3}$.
Therefore, the above formula should be reformulated by adding the volume fraction
of the heavy nucleus ($u$) to the Wigner-Seitz cell for the numerical calculation:
\begin{equation}
\ebth  = u \, \varepsilon_{\text{th}}^{\text{dense}} +(1-u)\, \varepsilon_{\text{th}}^{\text{dilute}}\,.
\end{equation}
For the pressure, it is not necessary to distinguish $p_{\text{th}}^{\text{dense}}$ and
$p_{\text{th}}^{\text{dilute}}$ because $p_{\text{th}}^{\text{dense}} = p_{\text{th}}^{\text{dilute}}$ 
in the equilibrium state.
The volume fraction $u$ should depend on the density and temperature in order to formally obtain the
hot dense matter EOS \cite{lim2017a}. 

In the present case the zero-temperature EOS is generated from
the liquid drop model \cite{lim2017a} to describe nuclear clustering, while at the specific values of temperature 
$T=5$, $10$, $20$, $40$, $80$\,MeV we assume uniform nuclear matter. 
Note that at finite temperature
nuclei can be formed surrounded by unbound neutrons and protons (a situation we will consider at the end
of this section). However, if the temperature is
greater than the critical temperature of the nuclear liquid-gas phase transition 
($T_c \simeq 20$\,MeV \cite{wellenhofer15,wellenhofer16,carbone18}), all nuclei dissociate and
our Fermi liquid theory calculation of the free energy $F$ is well justified.
Our Bayesian nuclear modeling does not provide the effective masses in a dense nuclear medium.
In order to include thermal effects, we
parameterize $\Gth$ according to different values of the proton and neutron effective masses according to
Eqs.\ (\ref{gamt}) and (\ref{eq:gth1}).
Thus, we take the effective masses of nucleons in symmetric nuclear matter $M_{\rm SNM}^*$
and pure neutron matter $M_{\rm PNM}^*$
at saturation density as parameters within the range of accepted values, while the density dependence
follows from Eq.\ \eqref{eq:effm1}.

In Fig.\ \ref{fig:gthdist} we show the resulting $\Gth$ contour plots as a function of the total baryon number density for labeled
values of the nucleon effective masses at nuclear matter saturation density in symmetric nuclear matter $M_{\rm SNM}^*$
and pure neutron matter $M_{\rm PNM}^*$.
Below twice saturation density, Bayesian modeling of the nuclear matter EOS indicates that
$\Gth$ converges to a tight range of numerical values for a given set of effective masses.
In the present approximation, the adiabatic index is strongly sensitive to the values of the effective 
masses once the matter composition (as a function of density) is fixed by the underlying cold
dense matter equation of state. The error bands shown in Fig.\ \ref{fig:gthdist} thus reflect
uncertainties in the proton fraction coming from the Bayesian modeling of the equation of state.
In particular, we see that at low temperatures ($T=5$\,MeV) 
the contour plots exhibit rather large uncertainties beyond 
four or five times saturation density when the ratio of the effective mass to bare mass is not 1.
This is caused by the fact that some nuclear EOSs from 
our modeling predict that the ground state energy of pure neutron matter is lower than that of
symmetric nuclear matter, so that the $\Gth$ would converge to a finite number:
\begin{equation}\label{eq:gthsimp}
\Gth(n) = 1 + \frac{2}{3} 
\left[ 1 - \frac{3n}{2M^*}\frac{\partial M^*}{\partial n}\right]
= \frac{5}{3} + \frac{\beta n}{1 + \beta n}\,.
\end{equation}
Thus the upper limit of $\Gth$ would become $\Gth (n\rightarrow \infty) = \frac{8}{3}$
which we can see in Fig.\ \ref{fig:gthdist}. In the case of $M_{n}^*/M=1$ and $M_{p}^*/M=1$ 
$(\beta =0)$, the value of $\Gth$ becomes $\frac{5}{3}$ as we can see in the bottom of the Fig.\ \ref{fig:gthdist}.

\begin{figure}[t]
	\centering
	\includegraphics[scale=0.58]{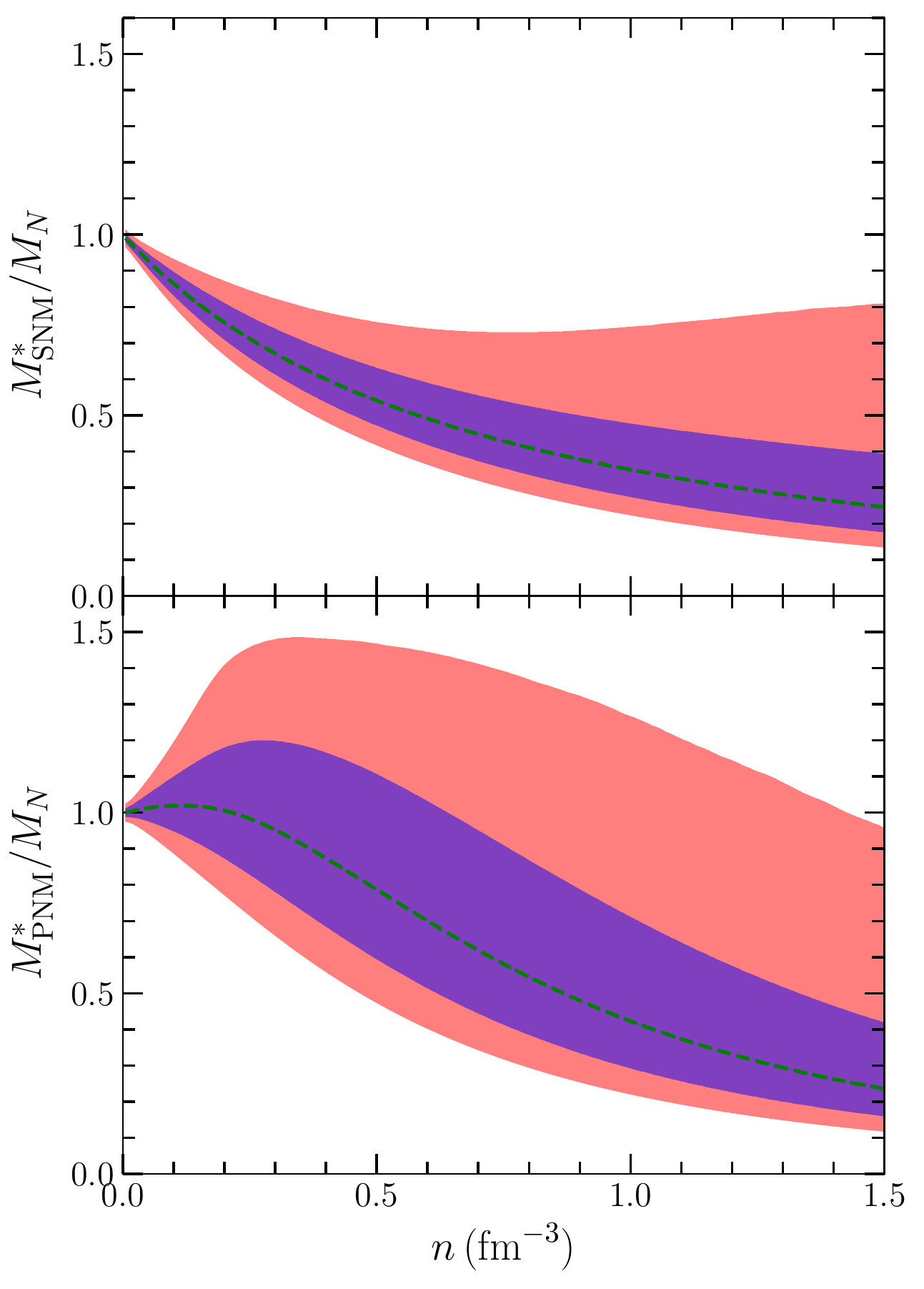}
	\caption{
	Uncertainties in the nucleon effective mass in symmetric nuclear matter (top) and pure neutron matter (bottom) 
	obtained from the extended Skyrme force models fitted to $\chi$EFT finite-temperature equations of state.}
	\label{fig:stateffm}
\end{figure}

We have seen that the parameterization for proton and neutron effective masses in Eq.\ \eqref{eq:effm1}
does not describe well the results from chiral effective field theory. A more general parameterization based 
on the extended form of the Skyrme functional may therefore provide a better approximation
for the thermal contributions to the pressure and energy density.
In our work, we have used $\gamma_1=\gamma_2=1$ in Eq.\ \eqref{lu}, thus the effective mass is a function of
$\beta_{L,U}$ and $\zeta_{L,U}$. The overall density dependence of the effective mass is then
described by the mean vector $\mathbf{x}=(\beta_L,\beta_U, \zeta_L, \zeta_U)$ of each coefficient	
\begin{equation}
\langle \mathbf{x} \rangle = (-7.61, 71.00, 33.88, -20.39)
\end{equation}
and its covariance matrix
\begin{equation}
\mathrm{Cov}_{\mathrm{x}} =
\begin{pmatrix}
      203.45  &    -130.00  &     78.95  &    202.92 \\
      -130.00 &     329.22  &    -36.18  &   -202.27 \\
      78.95   &   -36.18    &  145.90    &   -121.46 \\
      202.92  &   -202.27   &  -121.46   &    748.91 \\
\end{pmatrix}.
\end{equation}
In Fig.\ \ref{fig:stateffm} we show the resulting $1 \sigma$ and $2\sigma$ uncertainty bands on the nucleon
effective mass in symmetric nuclear matter (top panel) and pure neutron matter (bottom panel) 
as a function of density in the extended Skyrme mean field models fitted to the hot and dense 
matter equations of state from chiral effective field theory. 
The green dashed curves represent the statistical average effective masses among 100,000 effective mass parameter sets.
The behavior of effective mass is eventually decreasing as the total 
baryon number density increases, which guarantees the stabilities of the effective masses.

\begin{figure*}[t]
\centering
\includegraphics[scale=0.6]{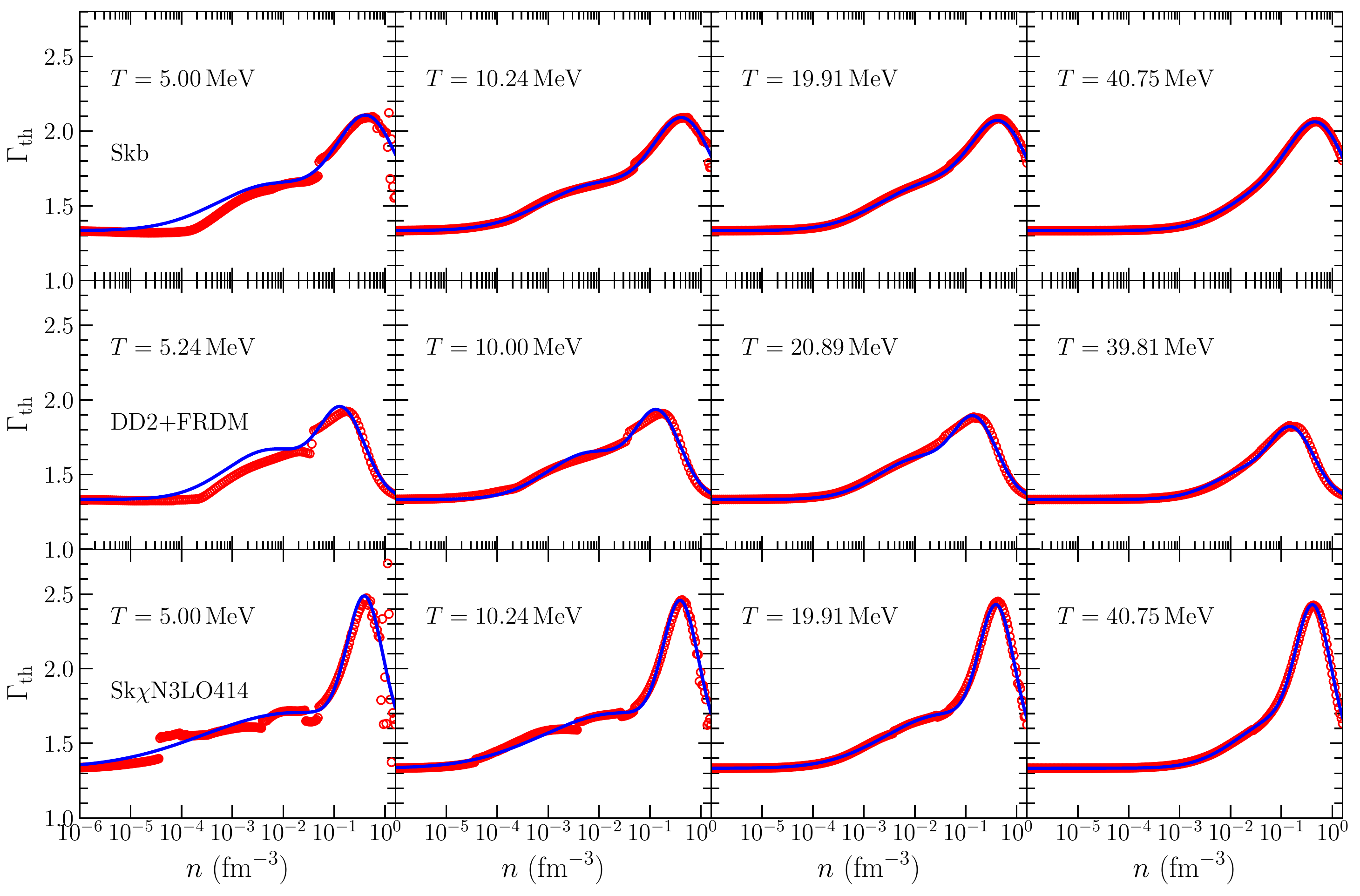}
\caption{
Density and temperature dependence of $\Gth$ from nuclear equations of state obtained from the 
mean field models (red circle) Skb, DD2-FRDM, and Sk$\chi$N3LO414 and the fitting function 
(blue curves) of Eq.\ (\ref{eq:gth}).}
\label{fig:gthzerocomp}
\end{figure*}

Next we consider microscopic modeling of thermal effects from finite-temperature quantum many-body
theory \cite{wellenhofer14,wellenhofer15}.
In this work, we follow the liquid drop model approach \cite{LSEOS} to construct the hot dense matter EOS
from Skyrme force models and $\chi$EFT.
The basic idea is to find the density of each species in the cell by minimizing the free energy
density for a given variable, in general within the three-dimensional space ($n$, $Y_p$, $T$).
Since the nuclear EOS depends on the baryon number density, temperature, and proton fraction,
$\Gth$ in general is a function of those variables as well. 
In the present study we consider the thermal index $\Gth$ for matter in beta-equilibrium, and therefore the 
proton fraction $Y_p$ is not an independent variable.
There is some ambiguity in how we define the proton fraction; whether to use its value at finite temperature
or zero temperature. 
We have computed $\Gth$ in both cases: (i) we use the same proton fraction for both finite temperature and 
zero temperature and (ii) we determine the proton fraction as a function of temperature for beta equilibrium matter. 
We have found that there is no significant difference between these two cases. 
Therefore, we have chosen to use the same proton fraction determined by the zero-temperature 
beta equilibrium condition also at finite temperature.
It is noteworthy that the baryon composition at high temperature becomes isospin-symmetric, 
since the magnitude of the neutron and proton chemical potentials is much greater than the electron chemical potential.

\begin{figure*}[t]
\centering
\includegraphics[scale=0.47]{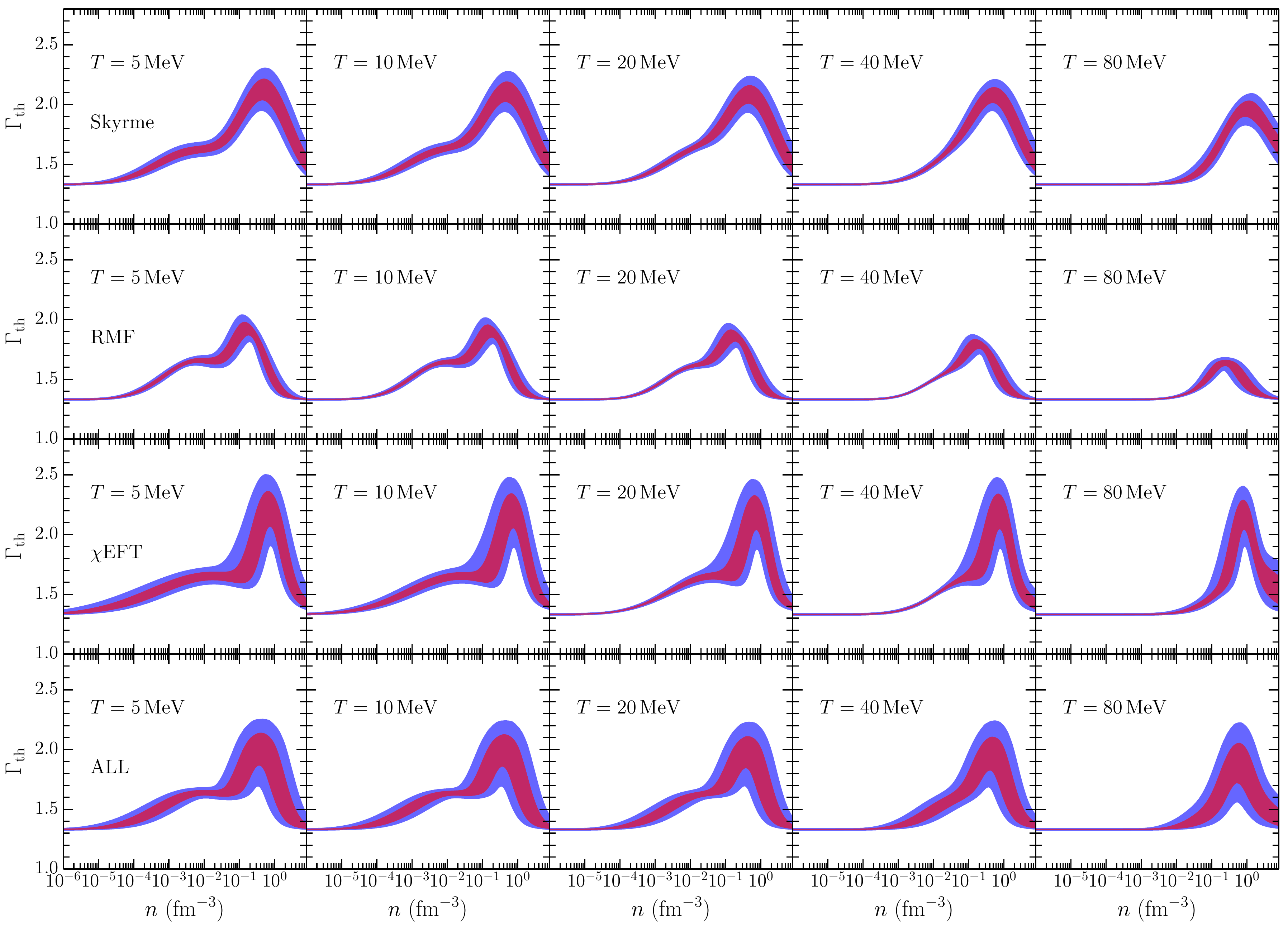}
\caption{Confident interval of $\pm 1\sigma$ and $\pm 2\sigma$ obtained from mean vector and covariance matrix
	for Skyrme, RMF, $\chi$EFT, and all combined models calculations.}
\label{fig:gth4x6credit}
\end{figure*}

By considering $\Gth$ from many EOSs, we propose a $\Gth$ fitting function:
\begin{equation}\label{eq:gth}
\Gth = \frac{4}{3} + a \exp[-b(\xi-\xi_a)^2] + c \exp[-d(\xi-\xi_b)^2]
\end{equation}
where $\xi=\log_{10}(n\,\mathrm{fm}^{3})$.
Note that we have introduced two Gaussian functions that reflect the behavior of $\Gth$ due to
two phase transition densities: one for neutron drip and the other
for the transition to uniform nuclear matter.
The constant $\frac{4}{3}$ corresponds to the low density ($n<10^{-5}\,\mathrm{fm}^{-3}$)
region where the dominant contribution to $\Gth$ comes from ultra relativistic electrons.
Since $\Gth$ depends on temperature, we assume that the parameters in Eq.\,\eqref{eq:gth}, 
$\mathbf{a} = (a, b, c, d, \xi_a,\xi_b)$
are linear functions of the temperature parameter $\mathcal{T}$:
\begin{equation}
a = a_0 + a_1 \mathcal{T}\,; \quad \mathcal{T} = \tanh\left(\frac{T}{100~\mathrm{MeV}}\right) \,.
\end{equation}
Thus, there are twelve parameters used to fit microscopic calculations of the thermal index as a function
of baryon number density and temperature.

In Fig.\ \ref{fig:gthzerocomp} we show results for $\Gth$ from a representative Skyrme model (Skb), 
RMF model (DD2+FRDM), and chiral EFT model (Sk$\chi$N3LO414) determined for beta-equilibrium matter.
For the relativistic mean field model DD2+FRDM (as well as those considered below), we use the EOS tables 
generated by Hempel {\it et al.} \cite{hempel10} to determine the values of $\Gth$.
In Fig.\ \ref{fig:gthzerocomp} the red circles represent the numerical 
calculations while the blue curves show the best-fit function of the form given in Eq.\ \eqref{eq:gth} fitted to
the EOS results for $T \ge 5$\,MeV. We see that in most cases the chosen fitting function can accurately reproduce the 
behavior of $\Gth$ for a wide range of densities and temperatures.
Overall, the adiabatic index from the microscopically constrained Sk$\chi$N3LO414 effective interaction 
exhibits a stronger peak (and a corresponding enhancement of thermal effects in astrophysical simulations) 
for densities in the range $1-2n_0$. Below we will find that this behavior is systematic across a wide range of 
chiral EFT and mean field models.

For several equations of state we find a discontinuity in $\Gth$ around 
$n=\frac{1}{2}n_0$ that comes from the phase transition from inhomogeneous nuclear matter to 
uniform nuclear matter. In these regions the analytical fitting function smooths out the discontinuity.
At low temperatures $T<5$\,MeV, nuclear clustering strengthens these discontinuities, leading to less reliable
fitting functions.
We reiterate that the proton fraction is not constant as a function of the baryon number density 
since the ground state of nuclear matter is determined by the chemical
equilibrium condition $\mu_n =\mu_p + \mu_e$. The effective mass is also not constant but rather has
a strong density dependence (see, e.g., Fig.\ \ref{fig:effm} above).
In our modeling, the maximum value of $\Gth$ occurs in the density region $n=0.4 - 0.7 \,\mathrm{fm}^{-3}$, 
which is significantly greater than the value $n=0.27\,\mathrm{fm}^{-3}$ found in Ref.\ \cite{Constantinou15}.
We note, however, that in Ref.\ \cite{Constantinou15} the density is found in the case of pure neutron matter.
In contrast, we consider beta equilibrium matter which has a finite proton fraction at all densities.

To obtain a representative set of $\mathbf{a}$ parameters, we use various nuclear EOS tables available in
the astrophysics simulation communities as well as our own EOS tables \cite{lim2018eos}.
In case of Skyrme force models, we obtain formulas for $\Gth$ using SLy4, SkI4, SkM$^*$, Ska, and Skb.
For RMF models we include DD2, FSG, IU-FSU, SFHo, SFHx, TM1, and TMA. 
Finally, for $\chi$EFT we use results from Skyrme interactions fitted to the finite-temperature equations of state
from the N2LO450, N2LO500, N3LO414, N3LO450, and N3LO500 chiral potentials.
We estimate theoretical uncertainties in $\mathbf{a}$ from the nuclear force models
by constructing the covariance matrix with elements
\begin{equation}
\mathbf{M}_{ij} = \langle (x_i-\langle x_i \rangle) (x_j-\langle x_j \rangle) \rangle\,.
\end{equation}
The covariance matrices for the three different classes of interactions as well as the combination of all
models are given by 

\begin{widetext}

\scriptsize
\begin{subequations}
\label{eq:covariance1}
\begin{alignat}{4}
\mathbf{v}_{\text{Sk}}  & = \kbordermatrix{
	& a_0 &  a_1 &  b_0 &  b_1 &  c_0 &  c_1 &  d_0 &  d_1 &  \xi_{a_0} &  \xi_{a_1}  &  \xi_{b_0}  &  \xi_{b_1} \\
& 2.74\text{e-1} 
& 7.96\text{e-2} 
& 4.32\text{e-1}
& 2.77\text{e-1}
& 7.82\text{e-1}
&    -6.71\text{e-1}
&     1.07\text{e+0}
&     7.70\text{e-1}
&    -2.47\text{e+0}
&     4.02\text{e+0}
&    -2.83\text{e-1}
&     2.72\text{e-1}    
}  \\
\mathbf{M}_{\text{Sk}} & = \kbordermatrix{
	& a_0 &  a_1 &  b_0 &  b_1 &  c_0 &  c_1 &  d_0 &  d_1 &  \xi_{a_0} &  \xi_{a_1}  &  \xi_{b_0}  &  \xi_{b_1}  \\
	a_0   &    1.01\text{e-3} 
	&  -2.50\text{e-3} 
	&  -3.66\text{e-4} 
	&   1.96\text{e-3} 
	&  -1.67\text{e-3} 
	&   3.47\text{e-3} 
	&   9.81\text{e-4} 
	&   2.65\text{e-3} 
	&   3.22\text{e-3} 
	&  -1.37\text{e-2} 
	&  -4.63\text{e-4} 
	&   1.85\text{e-3} \\
	a_1   &-2.50\text{e-3} 
	&   9.70\text{e-3} 
	&   1.78\text{e-4} 
	&  -3.06\text{e-3} 
	&   2.04\text{e-3} 
	&  -1.22\text{e-2} 
	&  -1.06\text{e-3} 
	&  -3.73\text{e-3} 
	&  -2.12\text{e-3} 
	&   2.60\text{e-2} 
	&  -3.58\text{e-3} 
	&  -1.11\text{e-2} \\
	b_0   &   -3.66\text{e-4} 
	&   1.78\text{e-4} 
	&   1.14\text{e-3} 
	&  -4.36\text{e-3} 
	&   4.20\text{e-3} 
	&  -2.32\text{e-3} 
	&  -2.47\text{e-3} 
	&  -6.19\text{e-3} 
	&  -9.62\text{e-3} 
	&   2.69\text{e-2} 
	&   4.56\text{e-3} 
	&   2.43\text{e-3} \\
	b_1   &1.96\text{e-3} 
	&  -3.06\text{e-3} 
	&  -4.36\text{e-3} 
	&   1.92\text{e-2} 
	&  -1.76\text{e-2} 
	&   2.11\text{e-2} 
	&   9.95\text{e-3} 
	&   2.65\text{e-2} 
	&   3.77\text{e-2} 
	&  -1.26\text{e-1} 
	&  -1.31\text{e-2} 
	&   2.79\text{e-3} \\
	c_0   &-1.67\text{e-3} 
	&   2.04\text{e-3} 
	&   4.20\text{e-3} 
	&  -1.76\text{e-2} 
	&   4.11\text{e-3} 
	&  -3.99\text{e-3} 
	&  -2.32\text{e-3} 
	&  -6.11\text{e-3} 
	&  -9.02\text{e-3} 
	&   2.83\text{e-2} 
	&   3.56\text{e-3} 
	&   3.82\text{e-4} \\ 
	c_1   & 3.47\text{e-3} 
	&  -1.22\text{e-2} 
	&  -2.32\text{e-3} 
	&   2.11\text{e-2} 
	&  -3.99\text{e-3} 
	&   1.65\text{e-2} 
	&   1.78\text{e-3} 
	&   6.54\text{e-3} 
	&   5.91\text{e-3} 
	&  -4.23\text{e-2} 
	&   3.12\text{e-3} 
	&   1.40\text{e-2} \\
	d_0   &  9.81\text{e-4} 
	&  -1.06\text{e-3} 
	&  -2.47\text{e-3} 
	&   9.95\text{e-3} 
	&  -2.32\text{e-3} 
	&   1.78\text{e-3} 
	&   6.04\text{e-3} 
	&   1.42\text{e-2} 
	&   2.07\text{e-2} 
	&  -6.19\text{e-2} 
	&  -8.93\text{e-3} 
	&  -3.68\text{e-3} \\
	d_1   &   2.65\text{e-3} 
	&  -3.73\text{e-3} 
	&  -6.19\text{e-3} 
	&   2.65\text{e-2} 
	&  -6.11\text{e-3} 
	&   6.54\text{e-3} 
	&   1.42\text{e-2} 
	&   3.68\text{e-2} 
	&   5.30\text{e-2} 
	&  -1.71\text{e-1} 
	&  -1.98\text{e-2} 
	&  -2.63\text{e-5} \\
	\xi_{a_0} & 3.22\text{e-3} 
	&  -2.12\text{e-3} 
	&  -9.62\text{e-3} 
	&   3.77\text{e-2} 
	&  -9.02\text{e-3} 
	&   5.91\text{e-3} 
	&   2.07\text{e-2} 
	&   5.30\text{e-2} 
	&   2.05\text{e-2} 
	&  -5.89\text{e-2} 
	&  -9.31\text{e-3} 
	&  -4.06\text{e-3} \\
	\xi_{a_1} & -1.37\text{e-2} 
	&   2.60\text{e-2} 
	&   2.69\text{e-2} 
	&  -1.26\text{e-1} 
	&   2.83\text{e-2} 
	&  -4.23\text{e-2} 
	&  -6.19\text{e-2} 
	&  -1.71\text{e-1} 
	&  -5.89\text{e-2} 
	&   2.13\text{e-1} 
	&   1.69\text{e-2} 
	&  -1.41\text{e-2} \\
	\xi_{b_0} &  -4.63\text{e-4} 
	&  -3.58\text{e-3} 
	&   4.56\text{e-3} 
	&  -1.31\text{e-2} 
	&   3.56\text{e-3} 
	&   3.12\text{e-3} 
	&  -8.93\text{e-3} 
	&  -1.98\text{e-2} 
	&  -9.31\text{e-3} 
	&   1.69\text{e-2} 
	&   6.52\text{e-3} 
	&   7.79\text{e-3} \\
	\xi_{b_1} & 1.85\text{e-3} 
	&  -1.11\text{e-2} 
	&   2.43\text{e-3} 
	&   2.79\text{e-3} 
	&   3.82\text{e-4} 
	&   1.40\text{e-2} 
	&  -3.68\text{e-3} 
	&  -2.63\text{e-5} 
	&  -4.06\text{e-3} 
	&  -1.41\text{e-2} 
	&   7.79\text{e-3} 
	&   1.65\text{e-2}
}
%%
%%
%%
%% \\
%%
%%
%%
\end{alignat}
\end{subequations}

\scriptsize
\begin{subequations}
\label{eq:covariance2}
\begin{alignat}{4}
\mathbf{v}_{\text{RMF}} & = \kbordermatrix{
	& a_0 &  a_1 &  b_0 &  b_1 &  c_0 &  c_1 &  d_0 &  d_1 &  \xi_{a_0} &  \xi_{a_1}  &  \xi_{b_0}  &  \xi_{b_1}\\
    &     3.43\text{e-1}   
    &    -3.38\text{e-1}   
    &     5.22\text{e-1}   
    &     8.29\text{e-1}   
    &     5.28\text{e-1}   
    &    -4.79\text{e-1}   
    &     2.69\text{e+0}
    &     9.99\text{e-1}   
    &    -2.31\text{e+0}   
    &     2.32\text{e+0}   
    &    -7.20\text{e-1}
    &     2.22\text{e-1}} 
\\
\mathbf{M}_{\text{RMF}} & = \kbordermatrix{
          & a_0 &  a_1 &  b_0 &  b_1 &  c_0 &  c_1 &  d_0 &  d_1 &  \xi_{a_0} &  \xi_{a_1}  &  \xi_{b_0}  &  \xi_{b_1}  \\
	a_0   &   4.90\text{e-4} 
	      &  -5.97\text{e-4} 
	      &  -2.30\text{e-4} 
	      &   8.91\text{e-4} 
	      &   2.65\text{e-4} 
	      &   3.39\text{e-5} 
	      &   6.48\text{e-3} 
	      &   5.00\text{e-3} 
	      &   2.00\text{e-3} 
	      &  -4.03\text{e-3} 
	      &   1.10\text{e-4} 
	      &  -9.73\text{e-4}\\
	a_1   &  -5.97\text{e-4} 
	      &   7.95\text{e-4} 
	      &   1.42\text{e-4} 
	      &  -8.12\text{e-4} 
	      &  -6.03\text{e-4} 
	      &   3.63\text{e-4} 
	      &  -9.69\text{e-3} 
	      &  -6.33\text{e-3} 
	      &  -2.40\text{e-3} 
	      &   5.33\text{e-3} 
	      &   4.55\text{e-4} 
	      &   1.82\text{e-3}\\
	b_0   &  -2.30\text{e-4} 
	      &   1.42\text{e-4} 
	      &   8.09\text{e-4} 
	      &  -1.87\text{e-3} 
	      &   1.03\text{e-3} 
	      &  -1.83\text{e-3} 
	      &   2.18\text{e-3} 
	      &  -4.21\text{e-3} 
	      &  -2.06\text{e-3} 
	      &   2.05\text{e-3} 
	      &  -2.54\text{e-3} 
	      &  -1.70\text{e-3}\\
	b_1   &   8.91\text{e-4} 
	      &  -8.12\text{e-4} 
	      &  -1.87\text{e-3} 
	      &   6.99\text{e-3} 
	      &  -1.37\text{e-3} 
	      &   3.10\text{e-3} 
	      &   6.88\text{e-3} 
	      &   2.38\text{e-2} 
	      &   6.92\text{e-3} 
	      &  -1.24\text{e-2} 
	      &   1.47\text{e-3} 
	      &   3.35\text{e-3}\\
	c_0   &   2.65\text{e-4} 
	      &  -6.03\text{e-4} 
	      &   1.03\text{e-3} 
	      &  -1.37\text{e-3} 
	      &   1.67\text{e-3} 
	      &  -1.99\text{e-3} 
	      &   1.35\text{e-2} 
	      &   1.39\text{e-3} 
	      &   9.12\text{e-4} 
	      &  -3.80\text{e-3} 
	      &  -1.98\text{e-3} 
	      &  -3.48\text{e-3}\\
	c_1   &   3.39\text{e-5} 
	      &   3.63\text{e-4} 
	      &  -1.83\text{e-3} 
	      &   3.10\text{e-3} 
	      &  -1.99\text{e-3} 
	      &   2.77\text{e-3} 
	      &  -1.25\text{e-2} 
	      &   2.85\text{e-4} 
	      &   4.49\text{e-4} 
	      &   2.22\text{e-3} 
	      &   3.55\text{e-3} 
	      &   3.83\text{e-3}\\
	d_0   &   6.48\text{e-3} 
	      &  -9.69\text{e-3} 
	      &   2.18\text{e-3} 
	      &   6.88\text{e-3} 
	      &   1.35\text{e-2} 
	      &  -1.25\text{e-2} 
	      &   3.05\text{e-1} 
	      &   1.05\text{e-1} 
	      &   5.15\text{e-2} 
	      &  -1.26\text{e-1} 
	      &  -1.88\text{e-2} 
	      &  -6.61\text{e-2}\\
	d_1   &   5.00\text{e-3} 
	      &  -6.33\text{e-3} 
	      &  -4.21\text{e-3} 
	      &   2.38\text{e-2} 
	      &   1.39\text{e-3} 
	      &   2.85\text{e-4} 
	      &   1.05\text{e-1} 
	      &   1.65\text{e-1} 
	      &   3.77\text{e-2} 
	      &  -9.13\text{e-2} 
	      &  -2.32\text{e-2} 
	      &  -1.22\text{e-2}\\
\xi_{a_0} &   2.00\text{e-3} 
          &  -2.40\text{e-3} 
          &  -2.06\text{e-3} 
          &   6.92\text{e-3} 
          &   9.12\text{e-4} 
          &   4.49\text{e-4} 
          &   5.15\text{e-2} 
          &   3.77\text{e-2} 
          &   8.30\text{e-3} 
          &  -1.60\text{e-2} 
          &   1.52\text{e-3} 
          &  -3.98\text{e-3}\\
\xi_{a_1} &  -4.03\text{e-3} 
          &   5.33\text{e-3} 
          &   2.05\text{e-3} 
          &  -1.24\text{e-2} 
          &  -3.80\text{e-3} 
          &   2.22\text{e-3} 
          &  -1.26\text{e-1} 
          &  -9.13\text{e-2} 
          &  -1.60\text{e-2} 
          &   3.62\text{e-2} 
          &   4.01\text{e-3} 
          &   1.13\text{e-2}\\
\xi_{b_0} &   1.10\text{e-4} 
          &   4.55\text{e-4} 
          &  -2.54\text{e-3} 
          &   1.47\text{e-3} 
          &  -1.98\text{e-3} 
          &   3.55\text{e-3} 
          &  -1.88\text{e-2} 
          &  -2.32\text{e-2} 
          &   1.52\text{e-3} 
          &   4.01\text{e-3} 
          &   9.70\text{e-3} 
          &   2.90\text{e-3}\\
\xi_{b_1} &  -9.73\text{e-4} 
          &   1.82\text{e-3} 
          &  -1.70\text{e-3} 
          &   3.35\text{e-3} 
          &  -3.48\text{e-3} 
          &   3.83\text{e-3} 
          &  -6.61\text{e-2} 
          &  -1.22\text{e-2} 
          &  -3.98\text{e-3} 
          &   1.13\text{e-2} 
          &   2.90\text{e-3} 
          &   9.14\text{e-3}
}
%%
%% \\
%%
%%
%%
\end{alignat}
\end{subequations}

\scriptsize
\begin{subequations}
\label{eq:covariance3}
\begin{alignat}{4}
\mathbf{v}_{\text{EFT}} & = \kbordermatrix{
	& a_0 &  a_1 &  b_0 &  b_1 &  c_0 &  c_1 &  d_0 &  d_1 &  \xi_{a_0} &  \xi_{a_1}  &  \xi_{b_0}  &  \xi_{b_1} \\
    &3.22\text{e-1}   
    &    -1.08\text{e-3}   
    &     1.60\text{e-1}   
    &     8.87\text{e-1}   
    &     7.62\text{e-1}   
    &    -2.65\text{e-1}   
    &     4.16\text{e+0}
    &     6.53\text{e+0}
    &    -1.93\text{e+0}
    &     2.99\text{e+0}
    &    -1.34\text{e-1}   
    &     3.29\text{e-2}
}  
\\
\mathbf{M}_{\chi\text{EFT}} & = \kbordermatrix{
	& a_0 &  a_1 &  b_0 &  b_1 &  c_0 &  c_1 &  d_0 &  d_1 &  \xi_{a_0} &  \xi_{a_1}  &  \xi_{b_0}  &  \xi_{b_1}  \\
a_0 &   1.58\text{e-3} 
    &  -4.21\text{e-3} 
    &  -2.10\text{e-3} 
    &   1.28\text{e-2} 
    &   5.97\text{e-4} 
    &   4.18\text{e-3} 
    &  -8.42\text{e-3} 
    &  -1.29\text{e-1} 
    &  -7.71\text{e-4} 
    &  -8.01\text{e-3} 
    &  -1.66\text{e-3} 
    &   8.11\text{e-4}\\
a_1 &  -4.21\text{e-3} 
    &   2.16\text{e-2} 
    &   5.73\text{e-3} 
    &  -3.68\text{e-2} 
    &   6.80\text{e-3} 
    &  -2.50\text{e-2} 
    &  -1.09\text{e-1} 
    &   2.61\text{e-1} 
    &   2.57\text{e-3} 
    &   5.23\text{e-2} 
    &  -1.46\text{e-3} 
    &  -1.66\text{e-3}\\
b_0 &  -2.10\text{e-3} 
    &   5.73\text{e-3} 
    &   2.93\text{e-3} 
    &  -1.68\text{e-2} 
    &  -1.42\text{e-3} 
    &  -6.07\text{e-3} 
    &   7.11\text{e-3} 
    &   1.63\text{e-1} 
    &   3.59\text{e-4} 
    &   1.09\text{e-2} 
    &   1.71\text{e-3} 
    &  -7.92\text{e-4}\\
b_1 &   1.28\text{e-2} 
    &  -3.68\text{e-2} 
    &  -1.68\text{e-2} 
    &   1.12\text{e-1} 
    &  -7.40\text{e-4} 
    &   3.79\text{e-2} 
    &  -3.84\text{e-3} 
    &  -1.10\text{e+0} 
    &  -3.70\text{e-3} 
    &  -1.00\text{e-1} 
    &  -2.28\text{e-2} 
    &   1.02\text{e-2}\\
c_0 &   5.97\text{e-4} 
    &   6.80\text{e-3} 
    &  -1.42\text{e-3} 
    &  -7.40\text{e-4} 
    &   1.27\text{e-2} 
    &  -8.39\text{e-3} 
    &  -9.72\text{e-2} 
    &  -4.60\text{e-2} 
    &   1.97\text{e-3} 
    &   3.12\text{e-2} 
    &  -2.27\text{e-4} 
    &  -2.48\text{e-3}\\
c_1 &   4.18\text{e-3} 
    &  -2.50\text{e-2} 
    &  -6.07\text{e-3} 
    &   3.79\text{e-2} 
    &  -8.39\text{e-3} 
    &   3.04\text{e-2} 
    &   1.61\text{e-1} 
    &  -2.28\text{e-1} 
    &  -6.66\text{e-4} 
    &  -6.57\text{e-2} 
    &   3.03\text{e-3} 
    &   1.40\text{e-3}\\
d_0 &  -8.42\text{e-3} 
    &  -1.09\text{e-1} 
    &   7.11\text{e-3} 
    &  -3.84\text{e-3} 
    &  -9.72\text{e-2} 
    &   1.61\text{e-1} 
    &   1.95\text{e+0} 
    &   1.65\text{e+0} 
    &   2.00\text{e-2} 
    &  -4.50\text{e-1} 
    &   4.97\text{e-2} 
    &  -7.51\text{e-3}\\
d_1 &  -1.29\text{e-1} 
    &   2.61\text{e-1} 
    &   1.63\text{e-1} 
    &  -1.10\text{e+0} 
    &  -4.60\text{e-2} 
    &  -2.28\text{e-1} 
    &   1.65\text{e+0} 
    &   1.22\text{e+1} 
    &   5.04\text{e-2} 
    &   6.67\text{e-1} 
    &   2.92\text{e-1} 
    &  -1.22\text{e-1}\\
\xi_{a_0} &  -7.71\text{e-4} 
          &   2.57\text{e-3} 
          &   3.59\text{e-4} 
          &  -3.70\text{e-3} 
          &   1.97\text{e-3} 
          &  -6.66\text{e-4} 
          &   2.00\text{e-2} 
          &   5.04\text{e-2} 
          &   5.01\text{e-3} 
          &  -6.50\text{e-3} 
          &  -2.20\text{e-3} 
          &   1.11\text{e-4}\\
\xi_{a_1} &  -8.01\text{e-3} 
          &   5.23\text{e-2} 
          &   1.09\text{e-2} 
          &  -1.00\text{e-1} 
          &   3.12\text{e-2} 
          &  -6.57\text{e-2} 
          &  -4.50\text{e-1} 
          &   6.67\text{e-1} 
          &  -6.50\text{e-3} 
          &   2.25\text{e-1} 
          &   2.56\text{e-2} 
          &  -1.47\text{e-2}\\
\xi_{b_0} &  -1.66\text{e-3} 
          &  -1.46\text{e-3} 
          &   1.71\text{e-3} 
          &  -2.28\text{e-2} 
          &  -2.27\text{e-4} 
          &   3.03\text{e-3} 
          &   4.97\text{e-2} 
          &   2.92\text{e-1} 
          &  -2.20\text{e-3} 
          &   2.56\text{e-2} 
          &   1.91\text{e-2} 
          &  -6.24\text{e-3}\\
\xi_{b_1} &   8.11\text{e-4} 
          &  -1.66\text{e-3} 
          &  -7.92\text{e-4} 
          &   1.02\text{e-2} 
          &  -2.48\text{e-3} 
          &   1.40\text{e-3} 
          &  -7.51\text{e-3} 
          &  -1.22\text{e-1} 
          &   1.11\text{e-4} 
          &  -1.47\text{e-2} 
          &  -6.24\text{e-3} 
          &   2.88\text{e-3}
}
%%
%% \\
%%
%%
\end{alignat}
\end{subequations}

\scriptsize
\begin{subequations}
\label{eq:covariance4}
\begin{alignat}{4}
\mathbf{v}_{\text{All}} & = \kbordermatrix{
	& a_0 &  a_1 &  b_0 &  b_1 &  c_0 &  c_1 &  d_0 &  d_1 &  \xi_{a_0} &  \xi_{a_1}  &  \xi_{b_0}  &  \xi_{b_1}  \\
&     3.18\text{e-1} 
&    -1.07\text{e-1} 
&     3.64\text{e-1} 
&     7.13\text{e-1} 
&     6.79\text{e-1} 
&    -4.47\text{e-1} 
&     2.84\text{e+0} 
&     3.02\text{e+0}
&    -2.21\text{e+0} 
&     2.99\text{e+0} 
&    -3.91\text{e-1} 
&     1.64\text{e-1}
}  \\
\mathbf{M}_{\text{All}} & = \kbordermatrix{
	& a_0 &  a_1 &  b_0 &  b_1 &  c_0 &  c_1 &  d_0 &  d_1 &  \xi_{a_0} &  \xi_{a_1}  &  \xi_{b_0}  &  \xi_{b_1}  \\
a_0 &  5.03\text{e-4} 
&  -1.82\text{e-3} 
&  -6.10\text{e-5} 
&   7.91\text{e-4} 
&  -5.95\text{e-4} 
&   1.40\text{e-3} 
&   1.04\text{e-3} 
&  -2.89\text{e-3} 
&   8.03\text{e-4} 
&  -6.68\text{e-3} 
&  -1.25\text{e-3} 
&  -8.61\text{e-5}\\
a_1 & -1.82\text{e-3} 
    &   1.22\text{e-2} 
    &  -7.84\text{e-4} 
    &  -2.43\text{e-3} 
    &   6.52\text{e-3} 
    &  -4.30\text{e-3} 
    &  -4.43\text{e-3} 
    &   1.81\text{e-2} 
    &   2.11\text{e-3} 
    &   3.67\text{e-2} 
    &   1.10\text{e-2} 
    &  -1.99\text{e-3}\\
b_0 & -6.10\text{e-5} 
    &  -7.84\text{e-4} 
    &   1.50\text{e-3} 
    &  -1.43\text{e-3} 
    &  -3.41\text{e-3} 
    &  -6.22\text{e-3} 
    &  -7.04\text{e-3} 
    &  -1.77\text{e-2} 
    &  -8.82\text{e-3} 
    &  -2.66\text{e-3} 
    &  -9.37\text{e-3} 
    &   3.82\text{e-3}\\
b_1 &  7.91\text{e-4} 
    &  -2.43\text{e-3} 
    &  -1.43\text{e-3} 
    &   7.52\text{e-3} 
    &  -4.28\text{e-3} 
    &   1.58\text{e-2} 
    &   1.46\text{e-2} 
    &  -1.02\text{e-2} 
    &   1.27\text{e-2} 
    &  -5.82\text{e-2} 
    &  -6.23\text{e-3} 
    &  -2.68\text{e-3}\\
c_0 & -5.95\text{e-4} 
    &   6.52\text{e-3} 
    &  -3.41\text{e-3} 
    &  -4.28\text{e-3} 
    &   5.74\text{e-3} 
    &  -1.13\text{e-3} 
    &  -2.29\text{e-3} 
    &   7.03\text{e-3} 
    &   1.94\text{e-3} 
    &   2.23\text{e-2} 
    &   7.58\text{e-3} 
    &  -1.81\text{e-3}\\
c_1 &  1.40\text{e-3} 
    &  -4.30\text{e-3} 
    &  -6.22\text{e-3} 
    &   1.58\text{e-2} 
    &  -1.13\text{e-3} 
    &   1.20\text{e-2} 
    &   1.42\text{e-2} 
    &   1.38\text{e-2} 
    &   9.24\text{e-3} 
    &  -2.43\text{e-2} 
    &   4.88\text{e-3} 
    &  -2.86\text{e-3}\\
d_0 &  1.04\text{e-3} 
    &  -4.43\text{e-3} 
    &  -7.04\text{e-3} 
    &   1.46\text{e-2} 
    &  -2.29\text{e-3} 
    &   1.42\text{e-2} 
    &   1.41\text{e-1} 
    &   1.99\text{e-1} 
    &   7.50\text{e-2} 
    &  -1.87\text{e-1} 
    &   3.04\text{e-2} 
    &  -3.65\text{e-2}\\
d_1 & -2.89\text{e-3} 
    &   1.81\text{e-2} 
    &  -1.77\text{e-2} 
    &  -1.02\text{e-2} 
    &   7.03\text{e-3} 
    &   1.38\text{e-2} 
    &   1.99\text{e-1} 
    &   7.44\text{e-1} 
    &   1.66\text{e-1} 
    &   6.25\text{e-2} 
    &   1.76\text{e-1} 
    &  -9.02\text{e-2}\\
\xi_{a_0} &  8.03\text{e-4} 
          &   2.11\text{e-3} 
          &  -8.82\text{e-3} 
          &   1.27\text{e-2} 
          &   1.94\text{e-3} 
          &   9.24\text{e-3} 
          &   7.50\text{e-2} 
          &   1.66\text{e-1} 
          &   1.53\text{e-2} 
          &  -1.58\text{e-2} 
          &   7.78\text{e-3} 
          &  -6.43\text{e-3}\\ 
\xi_{a_1} & -6.68\text{e-3} 
          &   3.67\text{e-2} 
          &  -2.66\text{e-3} 
          &  -5.82\text{e-2} 
          &   2.23\text{e-2} 
          &  -2.43\text{e-2} 
          &  -1.87\text{e-1} 
          &   6.25\text{e-2} 
          &  -1.58\text{e-2} 
          &   1.58\text{e-1} 
          &   3.34\text{e-2} 
          &   1.47\text{e-3}\\
\xi_{b_0} & -1.25\text{e-3} 
          &   1.10\text{e-2} 
          &  -9.37\text{e-3} 
          &  -6.23\text{e-3} 
          &   7.58\text{e-3} 
          &   4.88\text{e-3} 
          &   3.04\text{e-2} 
          &   1.76\text{e-1} 
          &   7.78\text{e-3} 
          &   3.34\text{e-2} 
          &   2.15\text{e-2} 
          &  -4.67\text{e-3}\\
\xi_{b_1} & -8.61\text{e-5} 
          &  -1.99\text{e-3} 
          &   3.82\text{e-3} 
          &  -2.68\text{e-3} 
          &  -1.81\text{e-3} 
          &  -2.86\text{e-3} 
          &  -3.65\text{e-2} 
          &  -9.02\text{e-2} 
          &  -6.43\text{e-3} 
          &   1.47\text{e-3} 
          &  -4.67\text{e-3} 
          &   4.69\text{e-3}
}
\end{alignat}
\end{subequations}
\end{widetext}
Theses numerical results can be used to generate new $\Gth$ parameters within uncertainties originated from 
nuclear model differences. 

In Fig.\ \ref{fig:gth4x6credit} we show the $1\sigma$ and $2\sigma$ uncertainty bands obtained from the mean values
and covariance matrices in Eqs.\ \eqref{eq:covariance1}--\eqref{eq:covariance4}. 
We see that as the temperature increases, the first inflection 
point disappears and the maximum value of $\Gth$ decreases in all cases.
In contrast to the results shown in Fig.\ \ref{fig:gthdist},
none of the mean field models considered in this section gives rise to a supersoft
symmetry energy that would result in pure neutron matter as a ground state at some large density. 
Therefore, the value of the adiabatic index never reaches the limiting case $\Gth \rightarrow \frac{8}{3}$
as indicated in Eq.\,\eqref{eq:gthsimp} where the low temperature approximation is valid.
Compared with Skyrme and $\chi$EFT interactions, 
the confidence band for RMF models is rather small. This may indicate that the range of effective masses in the
RMF considered in this work is smaller than from Skyrme and $\chi$EFT interactions.

\section{Summary}
\label{sec:sum}
We have investigated the thermal index $\Gth$ of hot and dense nuclear matter from 
microscopic many-body calculations based on chiral two- and three-body
interactions as well as from Skyrme and relativistic mean field models commonly used in the literature. 
We find that $\Gth$ depends strongly on the density, temperature, and proton fraction.
In the Fermi liquid approximation, $\Gth$ depends essentially on the proton and neutron effective masses 
at and above nuclear saturation density. However, the simple formulas for the thermal energy density
and pressure are valid only in the low-temperature range (low compared with the Fermi temperature).
The nucleon effective mass itself is expected to depend 
sensitively on the temperature \cite{donati94}, and as a result simple Skyrme force models cannot 
reproduce perfectly the finite-temperature results from chiral effective field theory.

We have parameterized $\Gth$ in terms of the nucleon effective masses in symmetric nuclear
matter and pure neutron matter by 
subtracting the internal energy density and pressure 
of the cold nuclear matter EOS constructed in the liquid drop model technique
from the values in uniform nuclear matter at finite temperature.
From this method, where the cold dense matter equation of state is sampled from a Bayesian
posterior probability distribution constrained by nuclear theory and experiment, we have obtained statistical 
uncertainties on the value of $\Gth$. As predicted from the Fermi liquid theory description of nuclear matter,
$\Gth$ sensitively depends on proton and neutron effective masses.

Finally, we extracted $\Gth$ from realistic hot and dense matter equations of state based on several Skyrme force 
models, relativistic mean field models, and results from $\chi$EFT that include an accurate treatment of the
inhomogeneous mixed phase at low density and finite temperature. We considered the case of matter in beta equilibrium, 
from which we determined the proton fraction as a function of density.
The mean field models constrained by
microscopic chiral effective field theory were found to exhibit a consistently larger adiabatic index compared to traditional Skyrme 
force models and especially relativistic mean field models. 
From these results we have suggested a simple but accurate 
phenomenological formula for $\Gth$ written as a function of density and temperature.
By combining the different models, we obtained mean values for the $\Gth$ parameters and their
corresponding covariance matrices. This parametrization can be used to supplement a wide range of cold 
equations of state for use in core-collapse supernova or neutron star merger simulations.

\acknowledgments
We thank A. Bauswein for discussion.
Work supported by the National Science Foundation under grant 
No.\ PHY1652199.  Portions of this research
were conducted with the advanced computing resources provided by Texas A\&M High Performance Research Computing.
Y. Lim was also supported by the Max Planck Society and the Deutsche Forschungsgemeinschaft (DFG, German Research Foundation) -- Projektnummer 279384907 -- SFB 1245.

\bibliographystyle{apsrev4-1}
\bibliography{biblio4}

\end{document}